\documentclass[acmsmall]{acmart}

\usepackage{pdflscape}  
\usepackage{longtable}  
\usepackage{multirow}   
\usepackage{array}      
\usepackage{booktabs}   
\usepackage{geometry}   
\usepackage{xcolor}     
\usepackage{graphicx} 
\usepackage{booktabs}   
\usepackage{siunitx}    
\usepackage{threeparttable} 
\usepackage{makecell}   
\usepackage{adjustbox}   
\usepackage[utf8]{inputenc} 
\usepackage{tcolorbox} 
\usepackage{tabularx} 
\usepackage{enumitem} 
\usepackage{tcolorbox} 
\usepackage{verbatim} 
\tcbuselibrary{listings,breakable} 
\usepackage{listings} 

\pdfoutput=1

\newtcblisting{promptbox}[1]{%
  breakable,
  enhanced,
  colback=gray!3,
  colframe=black!40,
  title={#1},
  listing only,
  listing options={
    basicstyle=\ttfamily\footnotesize,
    columns=fullflexible,
    breaklines=true,
    breakatwhitespace=false,
    keepspaces=true,
    showstringspaces=false
  }
}

\newcommand{\dec}[1]{\textcolor{red}{(#1)}}
\newcommand{\inc}[1]{\textcolor{green!60!black}{(#1)}}
\newcommand{\same}[1]{(#1)}

\newcounter{num}
\newcommand{\finding}[1]{ 
  \begin{tcolorbox}[boxsep=2pt, left=3pt, right=3pt, top=2pt, bottom=2pt]
    \textbf{Finding \refstepcounter{num}\thenum}: #1
  \end{tcolorbox}
}

\AtBeginDocument{%
  }

\setcopyright{acmlicensed}
\copyrightyear{2018}
\acmYear{2018}
\acmDOI{XXXXXXX.XXXXXXX}
\acmConference[Conference acronym 'XX]{Make sure to enter the correct
  conference title from your rights confirmation email}{June 03--05,
  2018}{Woodstock, NY}
\acmISBN{978-1-4503-XXXX-X/2018/06}

\begin{document}

\title{Scaling Test-Driven Code Generation from Functions to Classes: An Empirical Study}

\author{Yunhao Liang}
\email{liangyunhao22@mails.ucas.ac.cn}
\orcid{0009-0002-8104-0903}
\affiliation{%
  \institution{Chengdu Institute of Computer Applications, Chinese Academy of Sciences and University of Chinese Academy of Sciences}
  \country{China}
}

\author{Ruixuan Ying}
\email{ying.ruixuan.s5@dc.tohoku.ac.jp}
\orcid{0009-0008-9597-599X}
\affiliation{%
  \institution{Institute of Multidisciplinary Research for Advanced Materials (IMRAM), Tohoku University}
  \country{Japan}
}

\affiliation{%
  \institution{Institute of Multidisciplinary Research for Advanced Materials (IMRAM), Tohoku University}
  \country{Japan}
}

\author{Zhe Cui}
\email{cuizhe@casit.com.cn}
\orcid{}
\affiliation{%
  \institution{Chengdu Institute of Computer Applications, Chinese Academy of Sciences and University of Chinese Academy of Sciences}
  \country{China}
}

\author{Shiwen Ni}
\email{sw.ni@siat.ac.cn}
\orcid{}
\affiliation{%
  \institution{Shenzhen Key Laboratory for High Performance Data Mining, Shenzhen Institutes of Advanced Technology, Chinese Academy of Sciences}
  and
  \institution{Artificial Intelligence Research Institute, Shenzhen University of Advanced Technology}
  \country{China}
}




\renewcommand{\shortauthors}{Trovato et al.}

\begin{abstract}
Test-driven development (TDD) has been adopted to improve Large Language Model (LLM)-based code generation by using tests as executable specifications. However, existing TDD-style code generation studies are largely limited to function-level tasks, leaving class-level synthesis—where multiple methods interact through shared state and call dependencies—underexplored. In this paper, we scale test-driven code generation from functions to classes via an iterative TDD framework. Our approach first analyzes intra-class method dependencies to derive a feasible generation schedule, and then incrementally implements each method under method-level public tests with reflection style execution feedback and bounded repair iterations.
To support test-driven generation and rigorous class-level evaluation, we construct ClassEval-TDD, a cleaned and standardized variant of ClassEval with consistent specifications, deterministic test environments, and complete method-level public tests. 
We conduct an empirical study across eight LLMs and compare against the strongest direct-generation baseline (the best of holistic, incremental, and compositional strategies). Our class-level TDD framework consistently improves class-level correctness by +12 to +26 absolute points and achieves up to 71\% fully correct classes, while requiring only a small number of repairs on average. 
These results demonstrate that test-driven generation can effectively scale beyond isolated functions and substantially improve class-level code generation reliability. All code and data are available at \url{https://anonymous.4open.science/r/ClassEval-TDD-C4C9/}
\end{abstract}



\received{20 February 2007}
\received[revised]{12 March 2009}
\received[accepted]{5 June 2009}

\maketitle

\section{Introduction}
Large Language Models (LLMs) have demonstrated strong capability in generating executable code from natural language. This progress has stimulated growing interest in applying LLMs to practical software engineering workflows, including automated implementation, bug fixing, and test generation \cite{xia2023automated,yang2024evaluation,yuan2024evaluating,lee2025unidebugger}. However, despite their impressive fluency, LLM-generated code still frequently fails to satisfy functional requirements, especially when tasks involve non-trivial program structure, edge cases, or implicit constraints \cite{liu2023your}. In software testing and analysis, a natural way to address this reliability gap is to treat tests as executable specifications and use them to guide and validate code generation \cite{qi2015analysis,weimer2009automatically}.

Test-Driven Development (TDD) \cite{tdd} provides a principled paradigm for incrementally building correct software: developers write tests first, implement a small unit of functionality, and iteratively repair the implementation until all tests pass. Recent research has explored execution-guided and test-driven code generation with LLMs, showing that provding public tests during generation, running tests and feeding back failures can significantly improve correctness compared to one-shot generation \cite{mathews2024test,gehring2024rlef,liang2025recode}. However, existing studies and benchmarks largely focus on function-level generation, where each task is isolated and the generated code is evaluated independently \cite{humaneval,mbpp}. 
In contrast, real-world software is often organized around classes and object-oriented abstractions, where multiple methods interact through shared state, call dependencies, and class-level invariants \cite{classeval}. Scaling test-driven generation from methods to classes therefore introduces new challenges: class methods are interdependent, a method may call other methods to complete its functionality. In this setting, method implementation must follow a dependency order to avoid unresolved references or incomplete state. Moreover, failures may propagate across methods due to shared state or incorrect interactions, making repair more complex than simply fixing isolated functions.

In this paper we scale test-driven code generation from functions to classes and propose a dependency-aware class-level TDD framework that decomposes class generation into a sequence of method-scoped TDD loops. Our approach first analyzes intra-class method dependencies and predicts a feasible generation schedule order. It then implements each method incrementally following the schedule with a TDD loop: for each method, it generates an implementation with method-level public tests as guidance and executable specifications. If tests fail, we use a reflection-style repair mechanism inspired by Reflexion \cite{shinn2023reflexion}: it first reflects on the failure to identify the root cause, then generates a targeted patch to fix the issue rather than directly patching code from raw error messages \cite{shinn2023reflexion,madaan2023self,chen2025revisit}. This loop continues until all method tests pass or a bounded repair budget is exhausted.

A major obstacle to studying class-level TDD is the lack of a suitable benchmark with consistent specifications and reliable tests. We find that the widely used ClassEval \cite{classeval} benchmark contains multiple issues that confound evaluation, including incomplete or inconsistent docstrings, incorrect type annotations, mismatched class skeletons and implementations, non-unit private tests, redundant or duplicated tests, and tests with external environment dependencies (e.g., files, randomness, or time). Such issues can cause failures unrelated to model capability, thereby distorting conclusions about generation strategies and test-driven methods. To address this, we construct ClassEval-TDD, a repaired and standardized benchmark derived from ClassEval. ClassEval-TDD provides: (i) corrected class and method docstrings with unified reStructuredText style, (ii) aligned skeletons and method signatures consistent with intended behavior, (iii) deterministic and environment-safe private tests with automatic setup/teardown, and (iv) method-level public tests enabling TDD-style incremental synthesis.

We evaluate our approach on ClassEval-TDD using eight LLMs spanning multiple families and scales, comparing against strong direct-generation baselines including Holistic, Incremental, and Compositional strategies. Results show that dependency-aware class-level TDD consistently outperforms the best baseline across all models, improving \texttt{class\_success} by 12--26 absolute points and achieving up to 71\% fully correct classes. These gains are achieved with low repair overhead: under a bounded budget of at most three repair rounds per method, the average repair cost remains small (0.06--0.62 repairs per method), and most failures converge within a single additional correction step. Our dependency analysis further reveals that, despite high overall dependency F1, scheduling errors persist on a small set of systematically challenging tasks, where models favor plausible semantic workflows that conflict with true call dependencies.

In summary, this work makes three contributions: (1) ClassEval-TDD, a cleaned and reliable benchmark for studying class-level test-driven code generation; (2) a dependency-aware class-level TDD framework with reflection-based repair and (3) an empirical study across eight LLMs that quantifies correctness gains, repair cost, and systematic scheduling failure modes. Together, these results show that TDD is an effective and practical mechanism for improving reliability of LLM-based class generation, while also exposing the remaining composition and dependency challenges that must be addressed to scale test-driven generation to real-world software.

\section{Background and Motivation}
LLMs have made rapid progress in code generation, but reliability remains a key bottleneck for adoption in practical software engineering. While most existing evaluations focus on standalone function, real-world software is often organized around classes, where correctness depends on interactions among multiple methods.


\textbf{Why class-level generation is harder?}
Unlike isolated functions, class methods frequently share mutable state, call each other, and collectively maintain class invariants. As a result, class-level synthesis exhibits a composition gap: even if most methods are individually correct, the class may still fail due to cross-method inconsistencies, incorrect state transitions, or missing prerequisites.

\textbf{Why dependency-aware scheduling matters?}
Incremental generation is a natural approach for class synthesis, but it introduces an ordering problem: methods must be implemented in an order that respects intra-class call dependencies. If a prerequisite method is missing or incorrect, subsequent methods may fail to compile or behave incorrectly, leading to cascading errors and inefficient repair. Therefore, a key requirement for scalable class-level synthesis is to infer method dependencies and derive a feasible generation schedule.

\textbf{Why test-driven generation is promising?}
TDD treats tests as executable specifications and supports iterative refinement via execution feedback. Prior work has shown that test-driven and execution-guided generation can significantly improve function-level correctness. By extending this paradigm to the class level, we study dependency-aware class-level TDD: generating a complete class by implementing methods incrementally under tests, with bounded repair to control cost and avoid excessive trial-and-error.



\begin{figure*}[htbp]
  \centering
  \includegraphics[width=0.85\textwidth]{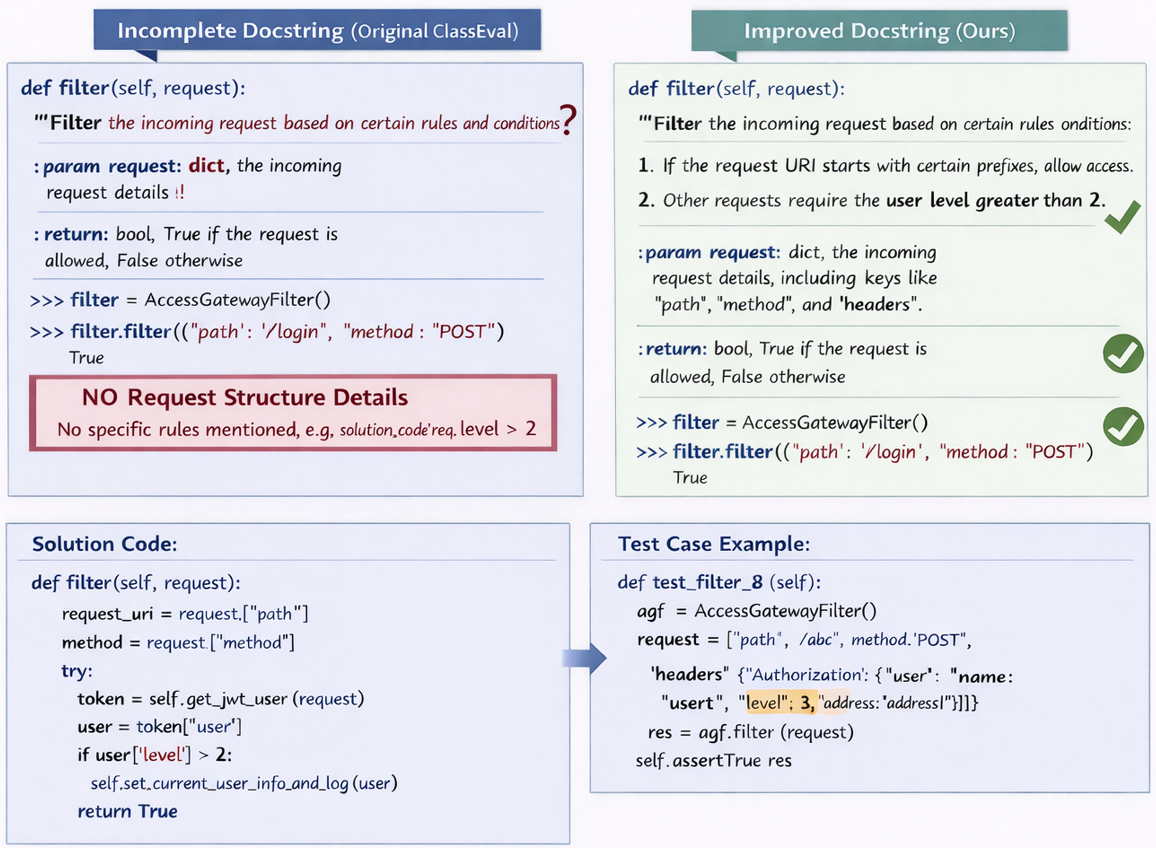} 
  \caption{An Example of Issue in ClassEval Benchmark}
  \Description{Example from ClassEval\_0. The original docstring omits the critical constraint that access is granted only when the user level is greater than 2, making the task underspecified and preventing correct implementations to pass tests.}
  \label{fig:docstring}
\end{figure*}

\section{Benchmark Construction: ClassEval-TDD}
A major obstacle to studying class-level test-driven generation is the lack of a reliable benchmark. We find that the widely used ClassEval contains multiple issues that confound evaluation, causing models to fail for reasons unrelated to code generation capability. To address this, we construct ClassEval-TDD, a cleaned and standardized benchmark derived from ClassEval.

\subsubsection{Limitations of ClassEval}
We observed three categories of issues in ClassEval that make rigorous evaluation difficult:

\textbf{Specification issues.}
Many tasks contain incomplete or inconsistent docstrings, incorrect type annotations, mismatched signatures, and skeleton/implementation inconsistencies. These issues make it impossible for models to infer critical constraints and write correct implementations.

\textbf{Test issues.}
Some tests are non-unit, redundant or duplicated, or inconsistent with tested methods. Such tests can lead to misleading failure signals and unreliable evaluation.

\textbf{Reliability issues.}
Some tasks rely on external environments such as files, randomness, or time, without proper isolation. This can lead to non-deterministic behavior and persistent side effects that affect subsequent executions.

\subsection{Construction Goals}
ClassEval-TDD is designed to support reliable class-level TDD research with three goals:
(1) \textbf{Correctness}: specifications and tests must be consistent and aligned with intended behavior;
(2) \textbf{Reliability}: tests must be deterministic and environment-safe to ensure reproducible execution;
(3) \textbf{TDD support}: each method must have method-level public tests for generation guidance.

\subsection{Repair and Standardization Pipeline}
Starting from the original ClassEval tasks, we apply a systematic repair pipeline.

\textbf{Code and skeleton alignment:}
We remove unused imports, eliminate redundant or dead code, fix inconsistent skeletons and signatures, and ensure that the class skeleton aligns with the solution code and corresponding tests.

\textbf{Docstring normalization:}
We correct spelling and formatting issues, unify docstrings into a consistent reStructuredText style, and complete missing specifications (parameters, return values, and doctest examples). For container-typed inputs (e.g., \texttt{dict}/\texttt{list}/\texttt{set}), we add structured descriptions of expected element types and key-value schemas.

\textbf{Deterministic and isolated private tests:}
We refactor private tests to follow unit-testing principles and ensure determinism. For tasks involving files, randomness, or time, we introduce isolation mechanisms (e.g., temporary files and fixed seeds) and enforce automatic cleanup via \texttt{setUp}/\texttt{tearDown}.

\textbf{Method-level public tests:}
To support TDD, we additionally annotate method-level public tests. Each method is paired with a dedicated test class containing multiple test cases (typically four, 1-2 for trivial methods), providing executable guidance during generation and repair.

\subsection{Benchmark Summary}
ClassEval-TDD contains 100 class tasks and 412 methods in total. Among them, 55 classes contain at least one method dependency, and 84 methods have non-empty dependency sets. The method-level public tests achieve high coverage of the repaired implementations (median 99.0\%, mean 98.7\%), enabling systematic evaluation of class-level TDD under strong test supervision. And private tests acheive 100\% line coverage for all tasks, ensuring reliable assessment. And we list all issues and their fixes in the appendix \ref{appendix:issues}.

\section{Method}
In this section, we introduce our class-level test-driven code generation framework, which consists of three main components: (i) dependency analysis and scheduling, (ii) method-level TDD generation, and (iii) reflective repair mechanism.

\subsection{Problem Definition}
We study class code generation under a test-driven setting. Given a class skeleton with serval method docstrings and method-level public tests, our goal is to generate a complete class implementation that passes all public tests during generation and all private tests during evaluation.

\begin{figure*}[htbp]
  \centering
  \includegraphics[width=0.8\textwidth]{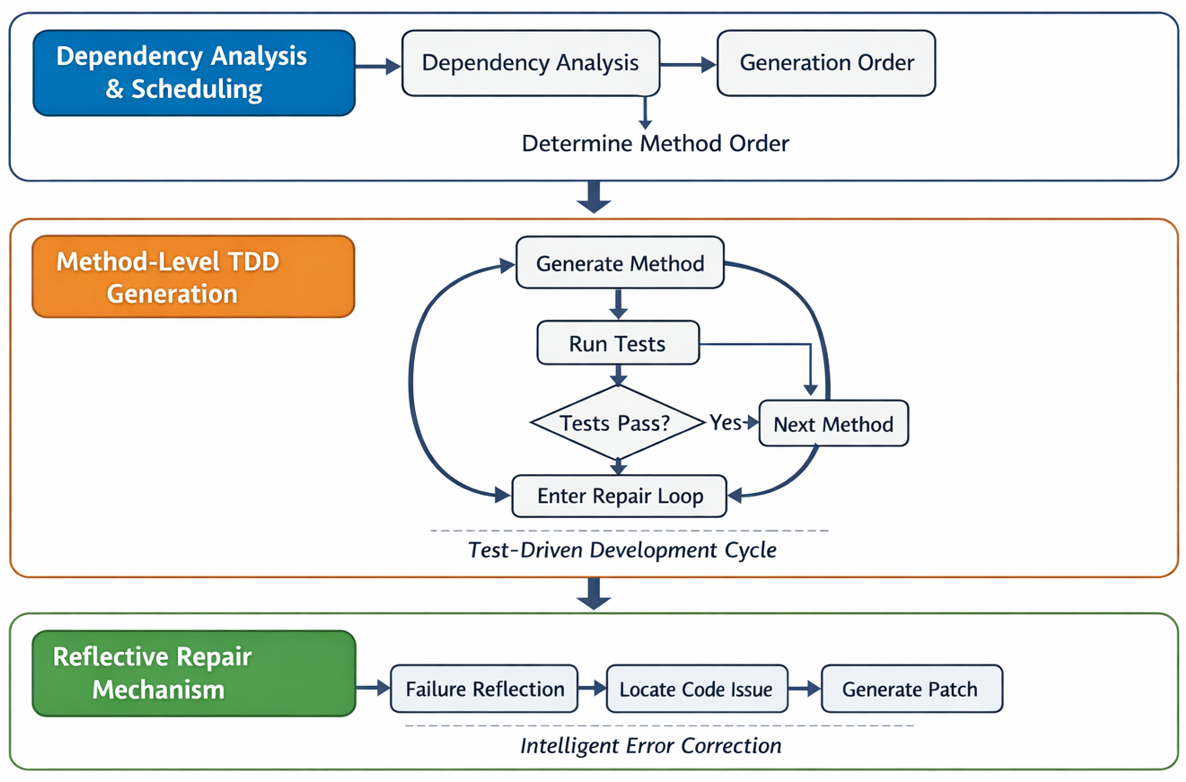} 
  \caption{Overview of our class-level TDD framework.}
  \Description{The framework consists of three main components: (1) Dependency analysis and scheduling; (2) Method-level TDD generation; (3) Reflective Repair Mechanism.}
  \label{fig:overview}
\end{figure*}

\subsection{Overview}
Figure \ref{fig:overview} summarizes our framework. It consists of three main components:
\begin{itemize}
    \item \textbf{Dependency analysis and scheduling.} We first analyze inter-method dependencies and predict a feasible schedule (generation order) that respects prerequisite relationships among methods.
    \item \textbf{Method-level TDD generation} Following the schedule order, we implement one method at a time using public tests as specifications and validation oracles. Each method is generated in a TDD loop: we generate the method, execute its public tests, and if any test fails, we enter a repair loop to fix the implementation until all tests pass or reach a repair budget.
    \item \textbf{Reflective Repair Mechanism:} Instead of directly rewriting code from raw error messages, we first prompt the LLM to reflect on the likely cause of the failure, identify the relevant code region, and then generate a minimal patch consistent with the intended specification.
\end{itemize}

\subsection{Dependency Analysis and Scheduling}
To enable class-level incremental test-driven generation, we first infer an implementation schedule that respects intra-class method dependencies. The key idea is to implement prerequisite (callee) methods before dependent (caller) methods, so that method-level tests can execute meaningful call chains without failing due to missing implementations.
In ClassEval-TDD, more than half of the classes contain at least one inter-method dependency, motivating the need for dependency-aware scheduling in class-level generation.

\subsubsection{Analysis Goal and Assumptions}
In class-level TDD, tests for a target method may exercise intra-class call chains (e.g., a public method invoking helper utilities). If a method under test calls another target method that has not yet been implemented, the execution may fail for structural reasons (e.g., missing attribute or undefined method) rather than incorrect functional logic. Such avoidable failures can mislead the repair loop and waste the limited repair budget. Therefore, we introduce a dependency-aware scheduling stage that aims to reduce missing-callee failures by implementing callees before callers whenever possible.

We adopt two assumptions consistent with our benchmark setting:
\begin{itemize}
    \item \textbf{Pre-implemented constructor.} The constructor \texttt{\_\_init\_\_} (if present) is already provided in the class skeleton. It is excluded from dependency analysis and does not appear in the generated method schedule.
    \item \textbf{Field availability.} Class fields initialized in \texttt{\_\_init\_\_} (e.g., \texttt{self.items}) are assumed to be available to all methods. Therefore, scheduling focuses on method-to-method dependencies rather than field initialization order.
\end{itemize}

\subsubsection{Dependency Analyze}
We prompt the LLM to act as a senior software architect and analyze the class skeleton together with the set of target methods $M$. It will analyzes each meethod's dependencies on other methods in $M$ and predict a feasible generation schedule. The LLM inspects each method $m_j \in M$ and determines whether it depends on any other method $m_i \in M$ based on two criteria:
\begin{itemize}
    \item \textbf{Direct call dependency.} The docstring of $m_j$ explicitly indicates that it calls $m_i$ (e.g., ``calls \texttt{m\_i} to ...'' or \texttt{self.m\_i(...)}).
    \item \textbf{Documented logical dependency.} The description of $m_j$ implies that its execution requires the behavior of $m_i$ as a prerequisite step (e.g., ``first format the input, then extract...''). We only treat such relations as dependencies when they are clearly stated in the specification, rather than inferred from undocumented assumptions.
\end{itemize}
Methods that do not depend on any other target methods are treated as \emph{standalone} and prioritized early in the schedule. This includes protected or helper methods (e.g., names starting with ``\_''), which are often reusable utilities and can support later methods once implemented.
Finally, the LLM outputs a generation schedule over $M$ that satisfies the predicted dependency constraints to guide subsequent TDD generation.

\subsection{Test-Driven Code Generation}
With the predicted method generation schedule, we proceed to implement each method in schedule using a TDD loop. For each target method $m$, we use its public method-level tests $T_{public}^{m}$ as executable specifications to guide generation and validate correctness.
\subsubsection{Context Construction}
For each target method $m$, we construct a prompt context that includes: the current partial class implementation (including all previously generated methods); the singnature and docstring of method $m$; the public tests $T_{public}^{m}$ for method $m$; and an explicit instruction to implement method $m$ according to its specification and pass all public tests.
The LLM is instructed to generate only the implementation of $m$. This design follows the TDD principle of implementing a small unit of functionality at a time, using tests as executable specifications.

\subsubsection{TDD Generation Loop}
We generate method $m$ using the constructed context and execute its public tests $T_{public}^{m}$. If all tests pass, we proceed to the next method in the schedule. If any test fails, we enter a repair loop to fix the implementation until all tests pass or a repair budget is exhausted.
After all methods in the schedule are generated, we obtain the final class implementation.

\subsection{Reflective Repair Mechanism}
Rather than directly patching code from raw failure output, ClassTDD adopts a structured Reflexion-style repair procedure that separates diagnosis from patch synthesis. For a failing method $m$, the model is asked to perform the following steps:
\begin{enumerate}
    \item \textbf{Failure analysis.} Identify the likely cause of the failure, such as syntax error, logical error, or dependency issue.
    \item \textbf{Code region identification.} Find out which part of the code is responsible for the failure (e.g., specific lines or blocks).
    \item \textbf{Repair suggestion.} Propose a plan about how the code should be modified to fix the issue.
    \item \textbf{Patch generation.} Generate a minimal code patch that implements the proposed fix while adhering to the method specification.
\end{enumerate}

\section{Experimental Setup}
\subsection{Research Questions}
\label{sec:rqs}
We structure our evaluation around the following research questions:

\begin{itemize}
    \item \textbf{RQ1 (Benchmark Impact).} How does repairing and standardizing ClassEval into ClassEval-TDD affect baseline generation performance?
    \item \textbf{RQ2 (Dependency Analysis).} How accurately can LLMs infer intra-class method dependencies and produce a feasible implementation schedule?
    \item \textbf{RQ3 (Effectiveness).} How effective is our class-level TDD framework?
\end{itemize} 

\subsection{Evaluated Models}
We evaluate a diverse set of LLMs spanning multiple families and model sizes, including deepseek-v3, gpt-oss-120B, qwen2.5-coder-7B/32B,, qwen3-coder-30B/480B, qwen3-235B and gemini3-flash.
All models are evaluated with greedy decoding (temperature=0) to ensure deterministic outputs.

\subsection{Baselines Generation Strategies}
As baselines, we include three generation strategies from ClassEval: \textbf{Holistic (H)}, which generates the entire class in one pass; \textbf{Incremental (I)}, which generates methods one at a time in and accumulates prior methods as context; and \textbf{Compositional (C)}, which generates each method independently using only the class skeleton and target method docstring as context.

\subsection{Class-level TDD Configuration}
We set a maximum of 3 repair attempts per method.

\subsection{Evaluation Metrics}
\subsubsection{Code Generation}
All evaluations use the private test suites in ClassEval-TDD.
Each method is associated with a private test class containing multiple unit test cases.
Besides method-level tests, each class task also includes a class-level test that evaluates inter-method interactions and overall class behavior.
We use the metrics defined in ClassEval for consistency and comparability.
\begin{itemize}
    \item \textbf{fun\_success:} the percentage of methods that pass all corresponding private test cases.
    \item \textbf{fun\_partial\_success:} the percentage of methods that at least pass one private test case but not fully correct.
    \item \textbf{class\_success:} the percentage of classes for which all methods are correct.
    \item \textbf{class\_partial\_success:} the percentage of classes that are not fully correct at the class level, but still exhibit partial correctness, meaning that at least one method has some passing tests.
\end{itemize}

\subsubsection{Dependency Analysis}
We also evaluate how well LLMs predict inter-method dependencies and valid schedules.

Method-level metrics:
\begin{itemize}
    \item \textbf{Exact Match:} the percentage of methods whose predicted dependency list exactly matches the ground-truth list.
    \item \textbf{Missing/Extra Dep:} the percentage of methods with at least one missing or extra (redundant) dependency.
    \item \textbf{Wrong Dep:} the percentage of methods with both missing and extra dependencies.
    \item \textbf{Precision, Recall, and F1:} macro-averaged over all dependency edges.
    \item \textbf{Class-level Accuracy:} the macro-average accuracy across all 100 class tasks.
    \item \textbf{Topological Order Violations:} the number of class tasks where the predicted method generation schedule is not a valid topological sort of the ground-truth dependency graph.
\end{itemize}

\section{Results and Analysis}

\begin{table}[htbp]
  \centering
  \caption{Comparison of Generation Strategies on ClassEval vs. ClassEval-TDD.}
  \label{tab:model_data_comparison}
  \resizebox{\textwidth}{!}{%
    \begin{tabular}{l l c c c c c c}
    \toprule
    \multirow{2}{*}{\textbf{Model}} & \multirow{2}{*}{\textbf{Metric}} & \multicolumn{3}{c}{\textbf{ClassEval}} & \multicolumn{3}{c}{\textbf{ClassEval-TDD}} \\
    \cmidrule(lr){3-5} \cmidrule(lr){6-8}
    & & \textbf{H} & \textbf{I} & \textbf{C} & \textbf{H} & \textbf{I} & \textbf{C} \\
    \midrule
    \multirow{4}{*}{deepseek-v3} 
    & Class Success          & 0.45 & 0.23 & 0.29 & 0.47 \textcolor{blue}{+0.02} & 0.46 \textcolor{blue}{+0.23} & 0.31 \textcolor{blue}{+0.02} \\
    & Class Partial & 0.73 & 0.40 & 0.49 & 0.73 & 0.66 \textcolor{blue}{+0.26} & 0.51 \textcolor{blue}{+0.02} \\
    & Function Success           & 0.7430 & 0.5837 & 0.6015 & 0.7695 \textcolor{blue}{+0.0265} & 0.7401 \textcolor{blue}{+0.1564} & 0.6602 \textcolor{blue}{+0.0587} \\
    & Function Partial   & 0.8685 & 0.7450 & 0.7808 & 0.9004 \textcolor{blue}{+0.0319} & 0.8611 \textcolor{blue}{+0.1161} & 0.7910 \textcolor{blue}{+0.0102} \\
    \midrule
    \multirow{4}{*}{gpt-oss-120B} 
    & Class Success          & 0.36 & 0.19 & 0.14 & 0.45 \textcolor{blue}{+0.09} & 0.29 \textcolor{blue}{+0.10} & 0.20 \textcolor{blue}{+0.06} \\
    & Class Partial & 0.71 & 0.42 & 0.35 & 0.72 \textcolor{blue}{+0.01} & 0.48 \textcolor{blue}{+0.06} & 0.35 \\
    & Function Success           & 0.6912 & 0.4621 & 0.4402 & 0.7344 \textcolor{blue}{+0.0432} & 0.5039 \textcolor{blue}{+0.0418} & 0.4570 \textcolor{blue}{+0.0168} \\
    & Function Partial   & 0.8685 & 0.6015 & 0.5617 & 0.8672 \textcolor{red}{-0.0013} & 0.5996 \textcolor{red}{-0.0019} & 0.5430 \textcolor{red}{-0.0187} \\
    \midrule
    \multirow{4}{*}{qwen2.5-coder-7B} 
    & Class Success          & 0.32 & 0.15 & 0.20 & 0.33 \textcolor{blue}{+0.01} & 0.16 \textcolor{blue}{+0.01} & 0.28 \textcolor{blue}{+0.08} \\
    & Class Partial & 0.54 & 0.30 & 0.43 & 0.54 & 0.29 \textcolor{red}{-0.01} & 0.49 \textcolor{blue}{+0.06} \\
    & Function Success           & 0.6434 & 0.4602 & 0.5857 & 0.6523 \textcolor{blue}{+0.0089} & 0.5242 \textcolor{blue}{+0.0640} & 0.6172 \textcolor{blue}{+0.0315} \\
    & Function Partial   & 0.8167 & 0.6195 & 0.7729 & 0.7988 \textcolor{red}{-0.0179} & 0.6819 \textcolor{blue}{+0.0624} & 0.7743 \textcolor{blue}{+0.0014} \\
    \midrule
    \multirow{4}{*}{qwen2.5-coder-32B} 
    & Class Success          & 0.37 & 0.28 & 0.22 & 0.43 \textcolor{blue}{+0.06} & 0.33 \textcolor{blue}{+0.05} & 0.31 \textcolor{blue}{+0.09} \\
    & Class Partial & 0.60 & 0.43 & 0.42 & 0.63 \textcolor{blue}{+0.03} & 0.50 \textcolor{blue}{+0.07} & 0.48 \textcolor{blue}{+0.06} \\
    & Function Success           & 0.6833 & 0.5558 & 0.5936 & 0.7129 \textcolor{blue}{+0.0296} & 0.6328 \textcolor{blue}{+0.0770} & 0.6191 \textcolor{blue}{+0.0255} \\
    & Function Partial   & 0.8526 & 0.7171 & 0.7530 & 0.8438 \textcolor{red}{-0.0088} & 0.7812 \textcolor{blue}{+0.0641} & 0.7656 \textcolor{blue}{+0.0126} \\
    \midrule
    \multirow{4}{*}{qwen3-coder-30B} 
    & Class Success          & 0.44 & 0.37 & 0.27 & 0.44 & 0.50 \textcolor{blue}{+0.13} & 0.33 \textcolor{blue}{+0.06} \\
    & Class Partial & 0.72 & 0.61 & 0.50 & 0.69 \textcolor{red}{-0.03} & 0.68 \textcolor{blue}{+0.07} & 0.52 \textcolor{blue}{+0.02} \\
    & Function Success           & 0.7291 & 0.6434 & 0.6096 & 0.7637 \textcolor{blue}{+0.0346} & 0.7422 \textcolor{blue}{+0.0988} & 0.6289 \textcolor{blue}{+0.0193} \\
    & Function Partial   & 0.8904 & 0.8028 & 0.7849 & 0.8906 \textcolor{blue}{+0.0002} & 0.8496 \textcolor{blue}{+0.0468} & 0.7656 \textcolor{red}{-0.0193} \\
    \midrule
    \multirow{4}{*}{qwen3-coder-480B} 
    & Class Success          & 0.47 & 0.41 & 0.31 & 0.55 \textcolor{blue}{+0.08} & 0.53 \textcolor{blue}{+0.12} & 0.46 \textcolor{blue}{+0.15} \\
    & Class Partial & 0.80 & 0.63 & 0.62 & 0.80 & 0.79 \textcolor{blue}{+0.16} & 0.71 \textcolor{blue}{+0.09} \\
    & Function Success           & 0.7968 & 0.7092 & 0.7092 & 0.8164 \textcolor{blue}{+0.0196} & 0.8047 \textcolor{blue}{+0.0955} & 0.7754 \textcolor{blue}{+0.0662} \\
    & Function Partial   & 0.9303 & 0.8606 & 0.8466 & 0.9297 \textcolor{red}{-0.0006} & 0.9297 \textcolor{blue}{+0.0691} & 0.9062 \textcolor{blue}{+0.0596} \\
    \midrule
    \multirow{4}{*}{qwen3-235B} 
    & Class Success          & 0.42 & 0.40 & 0.30 & 0.49 \textcolor{blue}{+0.07} & 0.48 \textcolor{blue}{+0.08} & 0.42 \textcolor{blue}{+0.12} \\
    & Class Partial & 0.74 & 0.69 & 0.64 & 0.74 & 0.72 \textcolor{blue}{+0.03} & 0.71 \textcolor{blue}{+0.07} \\
    & Function Success           & 0.7371 & 0.7331 & 0.6952 & 0.7773 \textcolor{blue}{+0.0402} & 0.7637 \textcolor{blue}{+0.0306} & 0.7598 \textcolor{blue}{+0.0646} \\
    & Function Partial   & 0.8745 & 0.8745 & 0.8566 & 0.9121 \textcolor{blue}{+0.0376} & 0.8926 \textcolor{blue}{+0.0181} & 0.9043 \textcolor{blue}{+0.0477} \\
    \midrule
    \multirow{4}{*}{gemini3-flash} 
    & Class Success          & 0.25 & 0.27 & 0.22 & 0.57 \textcolor{blue}{+0.32} & 0.59 \textcolor{blue}{+0.32} & 0.46 \textcolor{blue}{+0.24} \\
    & Class Partial & 0.30 & 0.37 & 0.35 & 0.82 \textcolor{blue}{+0.52} & 0.81 \textcolor{blue}{+0.44} & 0.63 \textcolor{blue}{+0.28} \\
    & Function Success           & 0.3088 & 0.5616 & 0.5319 & 0.8145 \textcolor{blue}{+0.5057} & 0.8242 \textcolor{blue}{+0.2626} & 0.7109 \textcolor{blue}{+0.1790} \\
    & Function Partial   & 0.3406 & 0.6545 & 0.6335 & 0.9062 \textcolor{blue}{+0.5656} & 0.9160 \textcolor{blue}{+0.2615} & 0.8242 \textcolor{blue}{+0.1907} \\
    \bottomrule
    \end{tabular}%
  }
\end{table}

\subsection{RQ1: Strategy Comparison on ClassEval-TDD}
ClassEval provides three representative class-level generation strategies: \textit{Holistic} (H), \textit{Incremental} (I), and \textit{Compositional} (C).
In this RQ, we investigate how repairing and standardizing ClassEval into ClassEval-TDD affects baseline generation performance across different models and strategies.
Table~\ref{tab:model_data_comparison} reports class-level and method-level performance under the three strategies on both datasets.
We summarize three key observations.

\textbf{Observation 1: Cleaning the benchmark consistently improves strategy performance, especially for unstable settings.}
Across models, ClassEval-TDD yields higher class\_success and function\_success under all three strategies, indicating that many failures on the original ClassEval are caused by benchmark noise rather than fundamental generation limitations.
The improvements can be substantial for certain models.
For example, \texttt{gemini3-flash} improves dramatically from 0.25 to 0.57 in class\_success under Holistic generation (+0.32), and from 0.27 to 0.59 under Incremental generation (+0.32).
Similarly, \texttt{gpt-oss-120B} improves from 0.36 to 0.45 under Holistic (+0.09) and from 0.19 to 0.29 under Incremental (+0.10).
These gains suggest that a reproducible benchmark with corrected specifications and deterministic tests is crucial for obtaining meaningful comparisons across generation strategies.
Across the eight evaluated models, ClassEval-TDD improves class\_success by an average of +0.08 under Holistic, +0.13 under Incremental, and +0.10 under Compositional generation, indicating that benchmark repair consistently strengthens class-level synthesis across strategies.
Moreover, the average performance gap between Holistic and Incremental shrinks from 0.098 on ClassEval to 0.049 on ClassEval-TDD, showing that cleaning substantially reduces the disadvantage of iterative generation.

\finding{Benchmark repair substantially improves class-level and method-level success across standard generation strategies, confirming that noisy specifications and non-reproducible tests can confound evaluation results.}

\textbf{Observation 2: Incremental generation becomes significantly more competitive on ClassEval-TDD.}
Incremental generation is particularly sensitive to benchmark noise because it accumulates previously generated methods as context.
On the original ClassEval, this accumulation can amplify inconsistencies (e.g., mismatched method signatures or ambiguous specifications), causing later methods to be generated under misleading context and resulting in compounding failures.
After cleaning, Incremental generation closes the gap with Holistic across most models, and even surpasses Holistic in some cases.
For instance, \texttt{qwen3-coder-30B} improves from 0.37 to 0.50 in class\_success under Incremental (+0.13), exceeding its Holistic performance (0.44).
For \texttt{deepseek-v3}, Incremental class\_success increases from 0.23 to 0.46 (+0.23), nearly matching Holistic (0.47).
Overall, these results show that once specifications and tests are standardized, Incremental generation becomes a strong baseline for class-level synthesis.

\finding{On ClassEval-TDD, Incremental generation becomes consistently more competitive and often approaches (or even surpasses) Holistic performance, indicating that strategy rankings on the original benchmark can be distorted by accumulated inconsistencies.}

\textbf{Observation 3: Compositional generation also benefits from benchmark repair, but remains limited by cross-method composition.}
Compositional generation constructs a class by generating methods independently and then assembling them, which reduces error propagation from earlier methods but sacrifices cross-method context.
On ClassEval-TDD, Compositional class\_success improves consistently (e.g., +0.09 for \texttt{qwen2.5-coder-32B}, +0.15 for \texttt{qwen3-coder-480B}, and +0.12 for \texttt{qwen3-235B}), suggesting that repaired docstrings and aligned skeletons substantially reduce spurious failures even for independent method synthesis.
Nevertheless, Compositional generation generally remains behind Holistic and Incremental, highlighting a class-level ``composition gap'': even when individual methods are plausible in isolation, they may not agree on shared state, naming conventions, or implicit invariants required for the class to function correctly as a whole.

\finding{Compositional generation also benefits from the cleaned benchmark, but typically lags behind due to cross-method composition challenges, motivating techniques that explicitly enforce class-level consistency.}

\textbf{Implications for class-level TDD:}
These findings highlight two practical implications.
First, reliable class-level evaluation requires benchmark standardization; otherwise, strategy comparisons may reflect dataset artifacts rather than true generation capability.
Second, the strong performance of Incremental generation on ClassEval-TDD suggests that iterative, context-aware synthesis is a promising foundation for class-level TDD, which further introduces test-guided repair and dependency-aware scheduling to improve correctness beyond direct generation.




\begin{table*}[htbp]
\centering
\caption{Class-level Method Dependency Analysis Results on ClassEval-TDD}
\label{tab:dep_analysis}
\resizebox{0.95\textwidth}{!}{%
\begin{tabular}{l c c c c c c c c c c}
\toprule
\multirow{2}{*}{\textbf{Model}} &
\multicolumn{7}{c}{\textbf{Method-level Metrics}} &
\multicolumn{3}{c}{\textbf{Class-level Metrics}} \\
\cmidrule(lr){2-8}\cmidrule(lr){9-11}
& 
\makecell{\textbf{Exact}\\\textbf{Match (\%)}} & 
\makecell{\textbf{Missing}\\\textbf{Deps (\%)}} & 
\makecell{\textbf{Extra}\\\textbf{Deps (\%)}} & 
\makecell{\textbf{Wrong}\\\textbf{Deps (\%)}} & 
\makecell{\textbf{Precision (\%)}} & 
\makecell{\textbf{Recall (\%)}} & 
\makecell{\textbf{F1 (\%)}} & 
\makecell{\textbf{Class-level}\\\textbf{Accuracy (\%)}} & 
\makecell{\textbf{Fully}\\\textbf{Correct Tasks}\\\textbf{(\#/100)}} & 
\makecell{\textbf{Topological}\\\textbf{Order Violations}\\\textbf{(\#/100)}} \\
\midrule
deepseek-v3 & 75.00 & 4.61 & 19.42 & 0.97 & 80.38 & 95.37 & 87.24 & 76.65 & 46 & 6 \\
gpt-oss-120B & 88.59 & 3.40 & 8.01 & 0.00 & 92.55 & 96.60 & 94.53 & 89.27 & 69 & 6 \\
Qwen2.5-coder-7B & 71.84 & 6.07 & 19.17 & 2.91 & 78.84 & 92.37 & 85.07 & 72.74 & 37 & 11 \\
qwen2.5-coder-32B & 84.47 & 3.88 & 10.92 & 0.73 & 88.96 & 95.89 & 92.29 & 84.43 & 60 & 7 \\
qwen3-coder-30B & 81.07 & 4.85 & 13.11 & 0.97 & 87.26 & 94.96 & 90.95 & 82.54 & 53 & 9 \\
qwen3-coder-480B & 82.77 & 4.85 & 11.65 & 0.73 & 88.35 & 95.37 & 91.72 & 84.12 & 60 & 4 \\
qwen3-235B & 87.86 & 5.34 & 6.55 & 0.24 & 93.75 & 94.90 & 94.32 & 88.70 & 66 & 6 \\
Gemini3-flash & 84.71 & 2.18 & 12.86 & 0.24 & 87.42 & 98.14 & 92.47 & 85.50 & 63 & 3 \\
\bottomrule
\end{tabular}
}
\end{table*}

\subsection{RQ2: Dependency Analysis and Method Scheduling}
Class-level synthesis is challenging because methods are not independent and a key component of our framework is to generate methods in a dependency-aware generation order. In this RQ, we evaluate the quality of dependency inference and the correctness of the resulting method schedules.

\subsubsection{Overall dependency prediction quality.}
Table~\ref{tab:dep_analysis} shows that LLMs can infer class-level method dependencies with consistently strong performance.
At the method level, \texttt{perfect\_match} ranges from 71.84\% to 88.59\%, and the average F1 score remains high (85.07\%--94.53\%).
A consistent pattern is that recall is typically higher than precision, suggesting that models tend to over-approximate dependencies rather than omit them.
At the class level, Task-level Accuracy reaches up to 89.27\%, and 37--69 tasks (out of 100) are fully correct depending on the model.

\finding{LLMs achieve consistently high recall (92.37\%--98.14\%) and strong F1 (85.07\%--94.53\%) for dependency inference, providing a viable signal for dependency-aware scheduling.}

Scheduling validity:
Beyond dependency analysis, we also examine whether the predicted method generation order is a valid topological sort of the ground-truth dependency graph.
As shown in Table~\ref{tab:dep_analysis}, topological order violations are relatively rare (3--11 tasks depending on the model), but non-negligible.
Notably, violations persist even for strong models (e.g., 6 tasks for gpt-oss-120B), suggesting that scheduling errors are not solely due to weak dependency prediction accuracy but may also stem from systematic semantic priors when ordering methods.

\finding{While dependency inference achieves high overall F1, LLMs still occasionally propose invalid method schedules (3--11 tasks), indicating that schedule construction remains a distinct failure mode beyond dependency-set prediction quality.}

\subsubsection{Where Dependency Inference Fails: Missing vs. Extra Dependencies}
To better understand failure modes, we examine the distribution of dependency errors.
Across models, the dominant error type is \texttt{extra\_deps}, where the predicted dependency set contains redundant methods not present in the ground truth.
This aligns with the high recall observed in Table~\ref{tab:dep_analysis} and suggests that models often adopt conservative guesses about call relations.
In contrast, \texttt{missing\_deps} occurs less frequently overall but is more harmful: missing a true prerequisite can invalidate the generation order and lead to downstream compilation or runtime failures when later methods assume missing functionality.

\finding{The primary dependency inference limitation is over-approximation (\texttt{extra\_deps}), while under-approximation (\texttt{missing\_deps}) is less frequent but more disruptive because it can invalidate schedules.}
\label{finding:rq2_error_modes}

\subsubsection{Dependency regimes: dependent vs. dependency-free methods}
While Table~\ref{tab:dep_analysis} reports aggregate statistics, dependency inference difficulty is not uniform across methods. We therefore stratify methods into two regimes: \texttt{With-deps} methods whose ground-truth dependency set is non-empty, and \texttt{No-deps} methods with empty dependency set. Table~\ref{tab:dep_stratified} reports error distributions for both groups.

\textbf{Dependency-free methods:}
For \texttt{No-deps} methods, \texttt{missing\_deps} and \texttt{wrong\_deps} are consistently 0\% across all models. This indicates that for stand alone methods, models almost never under-approximate. However, \texttt{extra\_deps} remains non-trivial (6.71\%--22.87\%), suggesting that models frequently hallucinate unnecessary prerequisites even when no dependency exists. Such redundant dependencies lower precision and may introduce avoidable ordering constraints.

\textbf{Dependency-bearing methods:}
In contrast, \texttt{With-deps} methods remain the main bottleneck. Missing dependencies become common (10.71\%--29.76\% \texttt{missing\_deps}), and \texttt{wrong\_deps} also increases (up to 14.29\%). These errors directly reduce recall and \texttt{exact\_match}, showing that the core challenge lies in identifying non-trivial call prerequisites rather than recognizing independence.

\finding{Dependency-free methods never miss dependencies (0\% \texttt{missing\_deps}) but frequently hallucinate redundant dependencies (\texttt{extra\_deps} up to 22.87\%), whereas dependent methods remain the main bottleneck with 10.71\%--29.76\% \texttt{missing\_deps}.}
\label{finding:rq2_dep_regime}

\begin{table*}[htbp]
\centering
\caption{Stratified Dependency Analysis by Dependency Regime on ClassEval-TDD.}
\label{tab:dep_stratified}
\setlength{\tabcolsep}{4pt}
\renewcommand{\arraystretch}{1.1}
\resizebox{\textwidth}{!}{%
\begin{tabular}{l|cccc|cccc}
\toprule
\multirow{2}{*}{\textbf{Model}} &
\multicolumn{4}{c|}{\textbf{With-deps Methods}} &
\multicolumn{4}{c}{\textbf{No-deps Methods}} \\
\cmidrule(lr){2-5}\cmidrule(lr){6-9}
& \makecell{\textbf{Exact}\\\textbf{Match (\%)}} 
& \makecell{\textbf{Missing}\\\textbf{Deps (\%)}} 
& \makecell{\textbf{Extra}\\\textbf{Deps (\%)}} 
& \makecell{\textbf{Wrong}\\\textbf{Deps (\%)}} 
& \makecell{\textbf{Exact}\\\textbf{Match (\%)}} 
& \makecell{\textbf{Missing}\\\textbf{Deps (\%)}} 
& \makecell{\textbf{Extra}\\\textbf{Deps (\%)}} 
& \makecell{\textbf{Wrong}\\\textbf{Deps (\%)}} \\
\midrule
deepseek-v3        & 66.67 & 22.62 & 5.95  & 4.76  & 77.13 & 0.00  & 22.87 & 0.00 \\
gpt-oss-120B       & 78.57 & 16.67 & 4.76  & 0.00  & 91.16 & 0.00  & 8.84  & 0.00 \\
qwen2.5-coder-7B   & 47.62 & 29.76 & 8.33  & 14.29 & 78.05 & 0.00  & 21.95 & 0.00 \\
qwen2.5-coder-32B  & 71.43 & 19.05 & 5.95  & 3.57  & 87.80 & 0.00  & 12.20 & 0.00 \\
qwen3-coder-30B    & 59.52 & 23.81 & 11.90 & 4.76  & 86.59 & 0.00  & 13.41 & 0.00 \\
qwen3-coder-480B   & 65.48 & 23.81 & 7.14  & 3.57  & 87.20 & 0.00  & 12.80 & 0.00 \\
qwen3-235B         & 66.67 & 26.19 & 5.95  & 1.19  & 93.29 & 0.00  & 6.71  & 0.00 \\
gemini3-flash      & 83.33 & 10.71 & 4.76  & 1.19  & 85.06 & 0.00  & 14.94 & 0.00 \\
\bottomrule
\end{tabular}%
}
\footnotesize
\texttt{Notes: With-deps} methods have at least one ground-truth dependency, while \texttt{No-deps} methods have none.
\end{table*}

\subsubsection{Error Analysis}
We further analyze three systematic topological-order failure cases (ClassEval\_44/61/94) that are consistently violated across models. Surprisingly, the ground-truth dependency graphs are extremely sparse, each containing only one or two dependency edges. However, models still output invalid orders due to strong semantic priors induced by method names (e.g., “extract-before-format”, “add/remove CRUD precedence”, “add-before-restock”), which conflict with the actual call dependencies. This indicates that LLMs tend to infer method generation orders based on natural language workflow rather than strict dependency constraints. We list all toploogical violations in the Appendix \ref{app:topo_violations}.

\finding{Topological violations are highly concentrated on a small set of tasks whose dependency graphs contain counter-intuitive edges or semantic-precondition dependencies. Models tend to follow semantic/workflow ordering, which systematically conflicts with such dependency constraints.}

\subsection{RQ3: Effectiveness of Dependency-Aware Class-Level TDD}
In this RQ, we evaluate the effectiveness of our class-level TDD framework in improving correctness over the best baseline generation strategies on ClassEval-TDD.

\subsubsection{Correctness Improvements}
Table~\ref{tab:tdd_vs_best_baseline} reports the performance of our TDD framework on ClassEval-TDD, compared to the best baseline strategy (max over Holistic/Incremental/Compositional) for each model. Overall, class-level TDD yields strong and consistent improvements across all evaluated LLMs.

At the class level, the best-performing models achieve up to 71\% \texttt{class\_success} (gpt-oss-120B and gemini3-flash). Several other strong models reach 68--70\% \texttt{class\_success} (deepseek-v3: 68\%, qwen3-coder-480B: 68\%, qwen3-235B: 70\%). Mid-sized models also achieve substantial performance (qwen2.5-coder-32B: 63\%, qwen3-coder-30B: 65\%). Even a 7B model reaches 46\% \texttt{class\_success}, indicating that test-driven code generation can significantly improve the usability of smaller models.

At the method level, correctness under TDD is consistently high for strong models: \texttt{fun\_success} exceeds 90\% for all large models, and \texttt{fun\_partial} further approaches 98--99\%. This indicates that the TDD loop is highly effective at driving individual methods to satisfy unit-level requirements.
However, class-level success remains notably lower than method-level success, revealing a persistent \emph{composition gap}: even when most methods are correct in isolation, integrating them into a fully correct class remains challenging due to shared state interactions and cross-method consistency constraints.


\finding{Dependency-aware class-level TDD consistently outperforms the strongest direct-generation baseline for every evaluated model, improving \texttt{class\_success} by +12 to +26 absolute points on ClassEval-TDD. Method-level correctness under TDD approaches saturation (90--92\% \texttt{fun\_success} for large models), yet class-level success remains substantially lower, highlighting a remaining composition gap driven by cross-method state and consistency constraints.}

\begin{table*}[htbp]
\centering
\caption{TDD vs. Best Baseline (max over H/I/C) on ClassEval-TDD (\%).}
\label{tab:tdd_vs_best_baseline}
\resizebox{\textwidth}{!}{%
\begin{tabular}{l c c c c c c c c}
\toprule
\multirow{2}{*}{\textbf{Model}} &
\multicolumn{2}{c}{\textbf{Class Success}} &
\multicolumn{2}{c}{\textbf{Class Partial}} &
\multicolumn{2}{c}{\textbf{Function Success}} &
\multicolumn{2}{c}{\textbf{Function Partial}} \\
\cmidrule(lr){2-3}\cmidrule(lr){4-5}\cmidrule(lr){6-7}\cmidrule(lr){8-9}
& \textbf{Best Baseline} & \textbf{TDD ($\Delta$)} 
& \textbf{Best Baseline} & \textbf{TDD ($\Delta$)}
& \textbf{Best Baseline} & \textbf{TDD ($\Delta$)}
& \textbf{Best Baseline} & \textbf{TDD ($\Delta$)} \\
\midrule
deepseek-v3 
& 47.00 (H) & 68.00 \textcolor{blue}{(+21.00)}
& 73.00 (H) & 93.00 \textcolor{blue}{(+20.00)}
& 76.95 (H) & 90.82 \textcolor{blue}{(+13.87)}
& 90.04 (H) & 98.44 \textcolor{blue}{(+8.40)} \\
\midrule
gpt-oss-120B 
& 45.00 (H) & 71.00 \textcolor{blue}{(+26.00)}
& 72.00 (H) & 93.00 \textcolor{blue}{(+21.00)}
& 73.44 (H) & 91.02 \textcolor{blue}{(+17.58)}
& 86.72 (H) & 98.63 \textcolor{blue}{(+11.91)} \\
\midrule
qwen2.5-coder-7B 
& 33.00 (H) & 46.00 \textcolor{blue}{(+13.00)}
& 54.00 (H) & 64.00 \textcolor{blue}{(+10.00)}
& 65.23 (C) & 74.22 \textcolor{blue}{(+8.99)}
& 79.88 (H) & 89.06 \textcolor{blue}{(+9.18)} \\
\midrule
qwen2.5-coder-32B 
& 43.00 (H) & 63.00 \textcolor{blue}{(+20.00)}
& 63.00 (H) & 82.00 \textcolor{blue}{(+19.00)}
& 71.29 (H) & 85.94 \textcolor{blue}{(+14.65)}
& 84.38 (H) & 95.31 \textcolor{blue}{(+10.93)} \\
\midrule
qwen3-coder-30B 
& 50.00 (I) & 65.00 \textcolor{blue}{(+15.00)}
& 69.00 (H) & 87.00 \textcolor{blue}{(+18.00)}
& 76.37 (H) & 87.89 \textcolor{blue}{(+11.52)}
& 89.06 (H) & 96.29 \textcolor{blue}{(+7.23)} \\
\midrule
qwen3-coder-480B 
& 55.00 (H) & 68.00 \textcolor{blue}{(+13.00)}
& 80.00 (H) & 97.00 \textcolor{blue}{(+17.00)}
& 81.64 (H) & 91.60 \textcolor{blue}{(+9.96)}
& 92.97 (I) & 99.22 \textcolor{blue}{(+6.25)} \\
\midrule
qwen3-235B 
& 49.00 (H) & 70.00 \textcolor{blue}{(+21.00)}
& 74.00 (H) & 92.00 \textcolor{blue}{(+18.00)}
& 77.73 (H) & 90.82 \textcolor{blue}{(+13.09)}
& 91.21 (H) & 98.24 \textcolor{blue}{(+7.03)} \\
\midrule
gemini3-flash 
& 59.00 (I) & 71.00 \textcolor{blue}{(+12.00)}
& 82.00 (H) & 92.00 \textcolor{blue}{(+10.00)}
& 82.42 (I) & 91.80 \textcolor{blue}{(+9.38)}
& 91.60 (I) & 98.05 \textcolor{blue}{(+6.45)} \\
\bottomrule
\end{tabular}%
}
\end{table*}

\subsubsection{Impact of intra-class dependencies.}
To better understand the remaining gap in class-level correctness, we further stratify tasks by whether methods/classes contain intra-class dependencies. Specifically, we separate (i) methods that depend on at least one other method and (ii) classes that contain at least one dependent method. Across all models, dependent methods are consistently harder: method-level \texttt{fun\_success} drops by 2--10 points compared to dependency-free methods. The effect is amplified at the class level, where classes with dependencies exhibit a substantially lower \texttt{class\_success} (8--16 points lower for strong models). Meanwhile, \texttt{fun\_partial} remains near-saturated for large models even on dependent methods, suggesting that failures are rarely catastrophic; instead, dependency-heavy classes tend to fail due to a small number of remaining inconsistencies in cross-method state, call ordering, or shared invariants. These results provide direct evidence that intra-class dependencies are a primary driver of the composition gap in class-level code generation.

\finding{Intra-class dependencies are a primary driver of class-level failures: across models, dependency-containing classes achieve 8--16 points lower \texttt{class\_success} than dependency-free classes under TDD.}

\begin{table}[htbp]
\centering
\caption{Impact of Intra-class Dependencies on TDD Performance (\%).}
\label{tab:dep_impact_tdd}
\setlength{\tabcolsep}{4pt}
\renewcommand{\arraystretch}{1.1}
\begin{tabular}{lcccc}
\toprule
\textbf{Model} &
\makecell{\textbf{Method}\\\textbf{Success}\\(with deps)} &
\makecell{\textbf{Method}\\\textbf{Success}\\(no deps)} &
\makecell{\textbf{Class}\\\textbf{Success}\\(with deps)} &
\makecell{\textbf{Class}\\\textbf{Success}\\(no deps)} \\
\midrule
deepseek-v3       & 88.1 & 90.6 & 63.6 & 73.3 \\
gpt-oss-120B      & 85.7 & 91.8 & 67.3 & 75.6 \\
qwen2.5-coder-7B  & 66.7 & 79.0 & 41.8 & 51.1 \\
qwen2.5-coder-32B & 82.1 & 87.5 & 58.2 & 68.9 \\
qwen3-coder-30B   & 79.8 & 89.9 & 58.2 & 73.3 \\
qwen3-coder-480B  & 86.9 & 90.9 & 63.6 & 71.1 \\
qwen3-235B        & 86.9 & 91.2 & 65.5 & 75.6 \\
gemini3-flash     & 86.9 & 92.7 & 63.6 & 80.0 \\
\bottomrule
\end{tabular}
\footnotesize
\\
\end{table}

\subsubsection{Repair Cost}
We further analyze the repair budget required by the TDD loop. The results are summarized in Table \ref{tab:tdd_repair_rounds}. Across strong models, the repair process is highly efficient, indicating that performance gains are not achieved via excessive trial-and-error but rather through targeted corrections on a small fraction of difficult methods.

At the method level, the average number of repair rounds per method is consistently low for large models: 0.06 for gpt-oss-120B, 0.12 for qwen3-coder-480B, 0.13 for qwen3-235B, 0.13 for gemini3-flash, and 0.15 for deepseek-v3. In terms of how often repairs are needed, only 15/412 methods require repair for gpt-oss-120B, while qwen3-480B, qwen3-235B, gemini3-flash, and deepseek-v3 require repairs for 30/412, 34/412, 32/412, and 34/412 methods, respectively. When considering only initially failing methods, the repair loop converges quickly, typically within 1.53–1.79 rounds for these strong models.

At the class level, repair costs remain similarly small. The average number of repair rounds per class is 0.07 for gpt-oss-120B, 0.12 for qwen3-480B, 0.14 for qwen3-235B and gemini3-flash, and 0.16 for deepseek-v3. Even among classes that require repair, convergence is fast: failed classes are fixed within 0.52–0.70 rounds on average for strong models, suggesting that many class-level failures can be resolved with at most one additional correction pass.

For smaller models, the repair cost increases but remains bounded. For qwen2.5-coder-7B, the average repair cost is 0.62 repairs/method, with 98/412 methods requiring repair, and failing methods converging in 2.60 rounds on average. These results demonstrate that class-level TDD remains practical under limited repair budgets, while providing substantial gains for both strong and compact models.

\finding{The correctness gains of class-level TDD come with modest overhead: under a strict budget of at most three repair rounds per method, strong models require fewer than 0.15 repairs per method on average, and most initially failing cases converge within fewer than two rounds.}

\begin{table*}[htbp]
\centering
\caption{Repair Cost of Class-level TDD}
\label{tab:tdd_repair_rounds}
\setlength{\tabcolsep}{4pt}
\renewcommand{\arraystretch}{1.15}
\begin{tabular*}{\textwidth}{l@{\extracolsep{\fill}}ccc|ccc}
\toprule
\multirow{2}{*}{\textbf{Model}} &
\multicolumn{3}{c}{\textbf{Method-level}} &
\multicolumn{3}{c}{\textbf{Class-level}} \\
\cmidrule(lr){2-4}\cmidrule(lr){5-7}
& \textbf{Avg} & \textbf{Need Repair} & \textbf{Fail Avg}
& \textbf{Avg} & \textbf{Need Repair} & \textbf{Fail Avg} \\
\midrule
deepseek-v3       & 0.15 & 34/412 & 1.79 & 0.16 & 22/100 & 0.70 \\
gpt-oss-120B      & 0.06 & 15/412 & 1.67 & 0.07 & 12/100 & 0.60 \\
qwen2.5-coder-7B  & 0.62 & 98/412 & 2.60 & 0.67 & 56/100 & 1.20 \\
qwen2.5-coder-32B & 0.26 & 46/412 & 2.33 & 0.28 & 33/100 & 0.85 \\
qwen3-coder-30B   & 0.25 & 50/412 & 2.06 & 0.27 & 33/100 & 0.83 \\
qwen3-coder-480B  & 0.12 & 30/412 & 1.60 & 0.12 & 21/100 & 0.58 \\
qwen3-235B        & 0.13 & 34/412 & 1.53 & 0.14 & 28/100 & 0.52 \\
gemini3-flash     & 0.13 & 32/412 & 1.62 & 0.14 & 23/100 & 0.62 \\
\bottomrule
\end{tabular*}
\footnotesize
\\
Avg = average repair rounds; Need Repair = number of methods or classes needing repair; Fail Avg = average rounds for initially failing methods or classes.
\end{table*}

\subsubsection{Effect of Reflection-based Repair}
To isolate the contribution of the reflection-based repair mechanism, we conduct an ablation study in which reflection is removed and failed methods are repaired by directly generating patches from raw error messages.
Table~\ref{tab:tdd_no_reflection} reports the resulting performance.

Overall, removing reflection leads to only a modest reduction in final correctness for strong models.
For large models such as gpt-oss-120B, qwen3-235B, qwen3-coder-480B, and gemini3-flash, class-level success remains within 1--3 absolute points of the full TDD framework.
This indicates that the main correctness gains primarily stem from test-driven incremental synthesis itself, rather than from reflection alone.

The impact of removing reflection becomes more noticeable for smaller and mid-sized models.
For example, qwen2.5-coder-7B drops from 46\% to 42\% in class-level success, and qwen2.5-coder-32B drops from 63\% to 62\%.
These results suggest that reflection-based repair provides additional guidance when model capacity is limited, helping stabilize repair behavior under bounded budgets.

The limited impact of reflection on final correctness can be explained by the overall repair statistics.
As shown in Table~\ref{tab:tdd_repair_rounds}, for strong models only a small fraction of methods and classes require any repair at all (e.g., 15--34 out of 412 methods, and 12--28 out of 100 classes).
Consequently, most methods pass their tests upon first generation, and only a minority of cases ever enter the repair loop.
In this regime, replacing reflection-based repair with direct error-driven patching affects only a small subset of difficult cases, which explains why the difference in final class-level correctness remains modest.

Taken together, these results show that reflection-based repair is not the primary driver of correctness gains.
Instead, its main contribution lies in improving robustness and efficiency on the relatively few hard cases that require repair, especially for smaller models.
This interpretation is further supported by the low average repair cost reported in Table~\ref{tab:tdd_repair_rounds}, which demonstrates that the TDD loop converges quickly without relying on excessive trial-and-error.

\finding{The correctness gains of class-level TDD largely arise from test-driven incremental synthesis itself, while reflection-based repair mainly improves robustness and efficiency on the small fraction of methods and classes that require repair, especially for weaker models under bounded repair budgets.}
\label{finding:rq3_reflection_effect}

\begin{table}[htbp]
\centering
\caption{TDD without Reflection-based Repair(\%).}
\label{tab:tdd_no_reflection}
\setlength{\tabcolsep}{6pt}
\renewcommand{\arraystretch}{1.15}
\begin{tabular}{l c c c c}
\toprule
\textbf{Model} &
\textbf{Class Success} &
\textbf{Class Partial} &
\textbf{Fun Success} &
\textbf{Fun Partial} \\
\midrule
deepseek-v3      
& 67.0 \dec{-1.0}
& 93.0 \same{+0.0}
& 90.43 \dec{-0.39}
& 98.44 \same{+0.0} \\

gpt-oss-120B     
& 70.0 \dec{-1.0}
& 93.0 \same{+0.0}
& 91.02 \same{+0.0}
& 98.63 \same{+0.0} \\

qwen2.5-coder-7B 
& 42.0 \dec{-4.0}
& 63.0 \dec{-1.0}
& 73.63 \dec{-0.59}
& 88.87 \dec{-0.19} \\

qwen2.5-coder-32B
& 62.0 \dec{-1.0}
& 81.0 \dec{-1.0}
& 84.96 \dec{-0.98}
& 95.12 \dec{-0.19} \\

qwen3-coder-30B  
& 67.0 \inc{+2.0}
& 88.0 \inc{+1.0}
& 89.45 \inc{+1.56}
& 97.07 \inc{+0.78} \\

qwen3-coder-480B 
& 65.0 \dec{-3.0}
& 94.0 \dec{-3.0}
& 89.84 \dec{-1.76}
& 98.44 \dec{-0.78} \\

qwen3-235B      
& 69.0 \dec{-1.0}
& 91.0 \dec{-1.0}
& 90.43 \dec{-0.39}
& 97.85 \dec{-0.39} \\

gemini3-flash   
& 71.0 \same{+0.0}
& 89.0 \dec{-3.0}
& 90.23 \dec{-1.57}
& 96.29 \dec{-1.76} \\
\bottomrule
\end{tabular}
\end{table}

\section{Discussion}
Our results demonstrate that class-level TDD can substantially improve correctness over strong non-TDD baselines on ClassEval-TDD. In this section, we discuss why the approach is effective, where it still fails, and what the results suggest for future class-level code generation and evaluation.

\subsection{Why Class-level TDD Works}
A key reason for the effectiveness of class-level TDD is that it provides executable constraints. Compared to one-shot class generation, our incremental TDD workflow provides two advantages.

First, method-level public tests provide direct behavioral feedback at a smaller granularity, allowing the model to correct mistakes before they propagate to the rest of the class. This is especially important for classes that mix simple standalone methods with more complex methods involving state transitions or edge cases. The consistently high method-level success rates under TDD indicate that many failures in baseline generation are not due to a lack of modeling capacity, but due to the absence of a feedback loop.

Second, the bounded repair loop enables the model to recover from imperfect initial generations with low overhead. Our repair-cost results show that most methods require no repair, and even failing methods typically converge within a small number of iterations. This suggests that test-driven refinement is not only effective but also cost-efficient for class-level synthesis.

\subsection{Dependency Scheduling: Useful but Imperfect}
More than half of the benchmark classes contain inter-method dependencies, which motivates dependency-aware scheduling. Our dependency analysis results show that LLMs can infer call-based dependencies with high accuracy, yet topological order violations still occur in a non-negligible number of tasks.

Interestingly, the most common error pattern is predicting \emph{extra} dependencies rather than missing true dependencies. This suggests that models often adopt conservative reasoning and over-approximate prerequisite relations. While over-approximation can still yield a feasible order, it may delay implementing some methods unnecessarily and reduce the effectiveness of incremental testing, especially when time or repair budgets are limited.

The observed order violations further indicate that, even when dependency edges are predicted correctly, models may fail to produce an order that satisfies all constraints. This gap highlights an opportunity for lightweight verification or post-processing (e.g., checking the predicted order against the predicted dependency graph and repairing the order if needed). In this paper, we intentionally keep the predicted order unchanged to isolate the behavior of prompt-based scheduling and avoid introducing additional algorithmic components that may confound the evaluation.

\subsection{Benchmark Quality as a First-class Concern}
Our benchmark comparison shows that repairing and standardizing ClassEval into ClassEval-TDD improves baseline performance for most models. This supports the view that benchmark noise can substantially distort conclusions in execution-guided generation settings.

In particular, class-level TDD is more sensitive to inconsistencies than one-shot generation: faulty tests, incomplete specifications, or non-deterministic behaviors can cause the repair loop to chase spurious failures and prevent convergence even when the implementation is reasonable. The improvements observed on ClassEval-TDD suggest that reliable evaluation of class-level generation requires not only stronger techniques but also cleaner benchmarks with deterministic and well-scoped unit tests.

\subsection{Implications for Class-level Code Generation}
Taken together, our findings suggest that scaling from method-level to class-level code generation benefits from combining three ingredients: (1) dependency-aware scheduling to reduce structural failures, (2) method-level executable specifications to guide incremental synthesis, and (3) localized repair to prevent non-local regressions. While stronger models achieve higher absolute success rates, the consistent improvements across model families indicate that the approach is broadly applicable and can serve as a practical technique for class-level generation.

Finally, we emphasize that partial success metrics remain important at the class level: even when a class is not fully correct, many methods can be synthesized correctly or partially correctly. This observation motivates future evaluation protocols that measure progress at multiple granularities and support realistic incremental development workflows.

\section{Related Work}

\subsection{Code Generation Benchmarks and Evaluation}
Most prior work on LLM-based code generation evaluates models on function-level benchmarks such as HumanEval and MBPP, where each task requires synthesizing a standalone function that satisfies a small test suite \cite{humaneval,mbpp,liu2023your}. These benchmarks have driven rapid progress, but their function-centric design under-represents challenges common in real-world software, including shared state, cross-method interactions, and class-level invariants.

To better capture object-oriented programming patterns, ClassEval \cite{classeval} proposes a class-level benchmark that evaluates complete class implementations under multiple generation strategies, including holistic, incremental, and compositional synthesis. Results on ClassEval show that class-level generation is substantially harder than function-level synthesis, as correctness depends on both individual methods and their integration. Our work builds on this motivation by studying test-driven generation at the class level and by providing a cleaned, reproducible benchmark to enable reliable evaluation of iterative synthesis techniques.

\subsection{Class-Level and Dependency-Aware Code Generation}
Many research improves code generation by introducing additional structure, such as decomposition, planning, or execution-based validation \cite{yao2022react,madaan2023self,chen2025revisit,zhong2024debug,chen2023teaching}. Decomposition-based approaches are a natural fit for class-level synthesis, but they require respecting inter-method dependencies arising from method calls, shared state, and initialization order.

Several works explore dependency-aware or planning-based generation \cite{jiang2024self,chen2025towards,zhou2022least}, but existing approaches typically focus on function-level tasks, rely on informal ordering heuristics, or do not evaluate dependency inference quality explicitly. In contrast, our work introduces an explicit dependency analysis and scheduling stage for class-level synthesis and evaluates it quantitatively using precision/recall-style metrics and topological-order validity. This analysis characterizes when dependency-aware incremental generation is effective and where dependency inference remains a bottleneck.

\subsection{Test-Driven Code Generation}
Test-driven development (TDD) treats tests as executable specifications and guides iterative implementation through execution feedback \cite{tdd,mathews2024test,piya2024llm4tdd,liu2025llm4tdg,tian2023test,zhangsynthesizing}. In research, test-driven ideas are most prominently explored in automated program repair (APR), where candidate patches are generated and validated against a test suite \cite{liang2025recode,ni2023lever}. More recently, execution feedback has been incorporated into LLM-based code generation, showing that iterative refinement can improve correctness over one-shot generation \cite{xia2023automated,gehring2024rlef,bouzenia2024repairagent,huang2023agentcoder,wang2023intervenor}.

However, Existing studies are largely limited to function-level tasks, where each target is synthesized in isolation. Our work studies test-driven synthesis at the class level, where correctness depends on interactions among multiple methods and shared state. Moreover, we separate public tests used during synthesis from hidden tests used for evaluation, reducing the risk of test overfitting. This combination of class-level scope, dependency-aware scheduling, and bounded test-driven repair distinguishes our approach from prior TDD-style code generation and APR research.




\section{Threats to Validity}

\paragraph{Internal validity.}
Our evaluation may be influenced by implementation details of the test-driven synthesis framework, such as the prompt design, repair budget, or runtime environment. To mitigate these risks, we use a fixed maximum of three repair rounds per method, apply consistent execution timeouts across models, and use greedy decoding to reduce variability. We also ensure determinism in the test environment by controlling randomness, file I/O, and time dependencies during execution.

\paragraph{Construct validity.}
We assess correctness using private private tests provided by ClassEval-TDD. While these tests offer high coverage and unit-level granularity, they may not fully capture all semantic behaviors or corner cases. Moreover, our dependency analysis focuses on method-level call relations and documented logical dependencies; it does not model implicit shared-state interactions or latent invariants, which may affect integration correctness. Nevertheless, the high alignment between method-level success and class-level behavior supports the relevance of our chosen metrics.

\paragraph{External validity.}
Our study is conducted on Python classes from ClassEval-TDD, which focuses on single-class tasks. The findings may not generalize to multi-class programs, complex object hierarchies, or languages with different object-oriented semantics (e.g., Java, C++). In addition, we assume the availability of method-level public tests, which may not be present in all real-world development settings. Future work should investigate class-level TDD in broader and less controlled environments, including multi-file systems and co-evolution of tests and code.

\section{Conclusion}
This paper investigated class-level code generation under a test-driven setting, motivated by the observation that generating correct classes is substantially harder than function-level synthesis due to stateful interactions and inter-method dependencies. We proposed a class-level TDD framework that synthesizes classes incrementally, generating one method at a time following a dependency-aware schedule and validating each method using public unit tests. When tests fail, our framework applies a reflection-before-patch repair loop under a bounded budget, enabling efficient convergence while preserving method-level locality during integration.

We conducted a comprehensive empirical study on ClassEval-TDD, a cleaned and reproducible benchmark derived from ClassEval with standardized specifications, deterministic tests, and hidden/private evaluation suites. Across a diverse set of LLMs, our results demonstrate that class-level TDD can substantially improve correctness under hidden tests, achieving 63–71\% class-level success for strong models while requiring only a small number of repairs on average (typically below 0.15 repair rounds per method). Our dependency analysis further indicates that LLMs can infer method dependencies with high recall but tend to over-approximate dependencies, suggesting that scheduling remains a meaningful lever for improving efficiency and stability. Overall, our findings provide the first systematic evidence that test-driven development is an effective and practical paradigm for class-level LLM-based code generation.


\section{Data Availability}
To support reproducibility and facilitate future research, we have released all artifacts associated with this paper, including the ClassEval-TDD benchmark, the evaluation scripts, and the implementation of our class-level TDD synthesis framework, in an anonymous public repository \url{https://anonymous.4open.science/r/ClassEval-TDD-C4C9/}. 
During the review period, we noly provide the anonymous repository link in the submission to enable artifact inspection while preserving author anonymity. After acceptance, we will update the repository to include author information and make it publicly accessible.


\bibliographystyle{ACM-Reference-Format}
\bibliography{ref}

\appendix
\label{appendix:issues}
\section{Appendix: Dataset Issues in ClassEval and Our Repairs for ClassEval-TDD}
This appendix documents (i) the issues we identified in the original ClassEval benchmark and (ii) the repairs and standardization steps we applied to construct ClassEval-TDD. We list issues exhaustively to improve transparency and reproducibility.

\subsection{Identified Issues in the Original ClassEval}

We found the following issues across class implementations, skeletons, docstrings, and private tests:

\subsubsection{Implementation-Level Issues}
\begin{enumerate}[label=\textbf{I\arabic*.}, leftmargin=*, itemsep=0.2em]
    \item \textbf{Unused imports}: invalid \texttt{import} statements where imported modules are never used.
    \item \textbf{Broken code or tests}: either the original implementation or the provided tests contain errors and fail when executed directly.
    \item \textbf{Defective implementations}: original code has functional defects and poor code quality.
    \item \textbf{Redundant code}: original implementation contains redundant logic that should be removed.
    \item \textbf{Commented-out statements}: implementation contains commented-out code fragments affecting clarity and correctness.
    \item \textbf{Empty \texttt{\_\_init\_\_}}: classes have no attributes, and the constructor is empty (\texttt{pass}); such constructors should be removed for consistency.
\end{enumerate}

\subsubsection{Skeleton/Metadata Mismatch and Missing Methods}
\begin{enumerate}[label=\textbf{S\arabic*.}, leftmargin=*, itemsep=0.2em]
    \item \textbf{Missing methods referenced in doctests}: doctests call methods that are absent in the skeleton or implementation (e.g., ClassEval-92).
    \item \textbf{Static method mismatch}: methods are called via the class name in doctests but are not declared as \texttt{@staticmethod} (e.g., ClassEval-62).
    \item \textbf{Doctest invocation without instantiation}: doctests call instance methods without creating an instance.
    \item \textbf{Implementation contains undocumented methods}: methods exist in implementation but are missing in the class skeleton or dataset metadata (e.g., ClassEval-25, ClassEval-49).
\end{enumerate}

\subsubsection{Docstring Specification Issues}
\begin{enumerate}[label=\textbf{D\arabic*.}, leftmargin=*, itemsep=0.2em]
    \item \textbf{Insufficient type specification for complex attributes}: for \texttt{dict}/\texttt{list} attributes not initialized via parameters, types are unclear or described informally (e.g., ClassEval-67 uses comments rather than a consistent convention).
    \item \textbf{Missing constructor docstrings}: constructors lack docstrings and do not distinguish constructor parameters from instance variables.
    \item \textbf{Non-standard naming}: method names are inconsistent or non-idiomatic (e.g., ClassEval-44).
    \item \textbf{Misspelled parameter names}: parameter name typos appear in signatures or docstrings.
    \item \textbf{Docstring name mismatches}: class/method names in docstrings do not match actual definitions.
    \item \textbf{Spelling/grammar errors}: docstrings contain spelling errors and English grammar issues.
    \item \textbf{Missing docstrings}: classes or methods have no docstrings at all.
    \item \textbf{Incomplete docstrings}: important conditions are missing, making correct generation impossible.
    \item \textbf{Doctest overlap}: docstring doctests overlap with private tests.
    \item \textbf{Missing return specification}: docstrings omit return value description, especially for methods operating on class state but returning values.
    \item \textbf{Inconsistent parameter/return fields}: some docstrings provide only types without descriptions, others only descriptions without types.
    \item \textbf{Incorrect type annotations}: type hints and/or docstring types are incorrect.
    \item \textbf{Invalid type names}: use of non-Python types such as \texttt{string} or \texttt{integer}.
    \item \textbf{Missing \texttt{:param}/\texttt{:return}}: docstrings omit parameters, returns, or doctests entirely.
    \item \textbf{Missing \texttt{:return} for \texttt{None}}: inconsistent treatment of no-return methods (sometimes \texttt{:return: None}, often omitted).
    \item \textbf{Inconsistent docstring style}: mixture of styles across methods/classes (ordering, punctuation, tense, capitalization).
    \item \textbf{Inconsistent verb tense}: docstrings begin with mixed verb forms (base vs third-person), inconsistent punctuation endings.
    \item \textbf{Missing static method annotation in docs}: static methods are not documented as static.
    \item \textbf{Under-specified complex inputs}: for complex input types (e.g., \texttt{dict} with required keys, or \texttt{object} parameters), docstrings omit critical structure while doctests imply the format (e.g., ClassEval-17, ClassEval-33).
\end{enumerate}

\subsubsection{Test Suite Issues}
\begin{enumerate}[label=\textbf{T\arabic*.}, leftmargin=*, itemsep=0.2em]
    \item \textbf{Duplicate private tests}: repeated test cases exist across the suite.
    \item \textbf{Redundant private tests}: multiple tests differ only superficially (e.g., \texttt{user1} vs \texttt{user2}) with little coverage gain.
    \item \textbf{Exact duplicates within a class}: fully identical test methods appear (e.g., ClassEval-75 \texttt{test\_remove\_item}).
    \item \textbf{External environment dependencies}: tests rely on external files, databases, randomness, or current time without proper isolation.
    \item \textbf{Missing environment cleanup}: failing tests do not clean up files/resources, affecting subsequent tests.
    \item \textbf{Missing setup/teardown}: tests do not use \texttt{setUp}/\texttt{tearDown} to prepare and clean environments.
    \item \textbf{Non-idiomatic naming}: private test class/method naming is inconsistent or mismatched with the tested method.
    \item \textbf{Tests call the wrong method}: a test labeled for method A calls method B (e.g., ClassEval-1 \texttt{test\_calculate\_sphere\_area} calls \texttt{calculate\_circle\_area}).
    \item \textbf{Non-unit tests}: tests for one method call other methods, violating unit-testing principles.
    \item \textbf{Test naming typos}: e.g., \texttt{delete} misspelled as \texttt{detele}.
    \item \textbf{Randomness/time not controlled}: tests involving randomness or time do not fix seeds or time.
    \item \textbf{Missing class-level tests}: some classes lack class-level test suites; naming conventions are inconsistent.
    \item \textbf{Inconsistent test class naming}: same class has multiple test-class naming patterns (abbreviations, casing differences).
    \item \textbf{Coverage gaps}: some methods lack dedicated method-level tests; number of test classes is smaller than number of methods.
\end{enumerate}

\subsection{Our Repairs and Standardization for ClassEval-TDD}
\label{app:dataset_repairs}

We applied the following repairs to construct ClassEval-TDD:

\subsubsection{Code and Skeleton Repairs}
\begin{enumerate}[label=\textbf{R\arabic*.}, leftmargin=*, itemsep=0.2em]
    \item \textbf{Diagnosed and fixed execution failures}: for samples where class code and tests fail to run, we determined whether failures stem from implementation or tests, and applied corresponding fixes.
    \item \textbf{Removed redundancy and fixed defects}: deleted redundant code and repaired incomplete/incorrect implementations.
    \item \textbf{Removed unused imports}: eliminated invalid or unused \texttt{import} statements.
    \item \textbf{Removed empty constructors}: deleted \texttt{\_\_init\_\_} methods that only contain \texttt{pass} when no initialization is required.
    \item \textbf{Added missing methods and aligned skeletons}: reconciled class skeletons, metadata, and implementations; added missing method stubs when doctests or implementations imply them.
\end{enumerate}

\subsubsection{Docstring and Type Standardization}
\begin{enumerate}[label=\textbf{R\arabic*.}, leftmargin=*, itemsep=0.2em, resume]
    \item \textbf{Added constructor docstrings}: for non-empty constructors, added docstrings and explicitly distinguished constructor parameters (\texttt{:param}) from instance variables (\texttt{:ivar}).
    \item \textbf{Fixed spelling and name inconsistencies}: corrected typos and mismatches across docstrings, parameter names, test class names, and test method names.
    \item \textbf{Completed and enriched specifications}: expanded incomplete docstrings using the intended behavior implied by implementations and tests, ensuring all constraints needed for correct generation are documented.
    \item \textbf{Enforced complete docstring structure}: ensured each method docstring includes (i) description, (ii) parameters, (iii) returns, and (iv) doctests where applicable.
    \item \textbf{Normalized type names}: replaced invalid \texttt{string}/\texttt{integer} with Python \texttt{str}/\texttt{int}.
    \item \textbf{Corrected type annotations}: fixed incorrect type hints and docstring types for parameters and return values.
    \item \textbf{Unified docstring style}: standardized to a consistent reStructuredText/Sphinx-like style with type-before-description ordering.
    \item \textbf{Unified verb tense}: normalized docstring opening verbs to base form for consistency.
    \item \textbf{Standardized \texttt{None} returns}: explicitly added \texttt{:return: None} for methods without return values, including short descriptions for state-mutating methods.
    \item \textbf{Detailed multi-case returns}: for methods with multiple return cases, documented each case under \texttt{:return:} with clear conditions.
    \item \textbf{Specified complex container types}: for \texttt{list}/\texttt{set}/\texttt{dict} parameters or attributes, documented element types and required key/value schemas (e.g., required keys and meanings).
    \item \textbf{Clarified static usage}: when doctests call methods via class name and methods do not depend on instance state, updated definitions and documentation to \texttt{@staticmethod} where appropriate.
\end{enumerate}

\subsubsection{Private Test Repairs and Determinization}
\begin{enumerate}[label=\textbf{R\arabic*.}, leftmargin=*, itemsep=0.2em, resume]
    \item \textbf{Environment-safe tests}: rewrote file/database-dependent tests to use temporary files and isolated environments.
    \item \textbf{Automated setup/cleanup}: added \texttt{setUp}/\texttt{tearDown} to prepare and clean environments to prevent cross-test interference.
    \item \textbf{Unified test naming}: standardized test class names and ensured they correspond to the tested methods/classes.
    \item \textbf{Added missing class-level tests}: created class-level test suites for classes lacking them.
    \item \textbf{Achieved full private-test coverage}: treated the repaired tests as private tests and ensured 100\% coverage for each sample.
    \item \textbf{Added method-level public tests}: for each method, manually created a corresponding public test class, typically containing four test methods (1--2 for simple methods), to enable TDD-style incremental generation.
    \item \textbf{Enforced unit-test principles}: refactored tests that depended on other class methods into unit tests that only invoke the target method, while preserving coverage.
    \item \textbf{Controlled randomness and time}: fixed random seeds for stochastic behaviors; normalized time formats and used fixed timestamps for time-dependent logic.
    \item \textbf{Removed duplication and redundancy}: eliminated duplicated test cases and reduced redundant variants that do not increase coverage.
\end{enumerate}

\subsection{Optional: Mapping Issues to Repairs}
\label{app:issue_repair_mapping}

For convenience, Table~\ref{tab:issue_repair_map} summarizes how our repairs address the identified issue categories.

\begin{table}[htbp]
\centering
\caption{High-level mapping from issue categories to repairs.}
\label{tab:issue_repair_map}
\setlength{\tabcolsep}{6pt}
\renewcommand{\arraystretch}{1.15}
\begin{tabular}{p{0.32\linewidth} p{0.60\linewidth}}
\toprule
\textbf{Issue Category} & \textbf{Representative Repairs} \\
\midrule
Implementation failures / redundancy & R1--R3 (execution fixes, remove redundancy, remove unused imports) \\
Skeleton/metadata mismatch & R5 (align skeletons, add missing methods) \\
Docstring incompleteness / inconsistency & R6--R15 (complete, standardize, fix names/types, enrich constraints) \\
Complex type underspecification & R16 (container schemas and element types) \\
Environment-dependent tests & R17--R18 (temporary files, setup/teardown) \\
Non-unit, redundant, duplicated tests & R20--R23 (unit-test refactoring, deduplication, determinization) \\
Coverage gaps / missing tests & R19--R21 (add class-level tests, add method-level public tests, ensure coverage) \\
\bottomrule
\end{tabular}
\end{table}

\section{Appendix: Topological Order Violations}
\label{app:topo_violations}

This appendix reports all tasks where the predicted method generation order is \emph{not} a valid topological ordering of the ground-truth dependency graph (i.e., it violates at least one dependency edge).
We denote a dependency edge as $A \rightarrow B$, meaning method $A$ \emph{depends on} method $B$ and thus $B$ should be generated \emph{before} $A$.

\subsection{Ground-truth Dependencies for Violated Tasks}
\label{app:gt_deps}

Table~\ref{tab:gt_deps_violated_tasks} summarizes the ground-truth method dependencies for all tasks that appear in topological order violations across our evaluated models.

\begin{table*}[htbp]
\centering
\small
\caption{Ground-truth method dependencies for tasks involved in topological order violations.}
\label{tab:gt_deps_violated_tasks}
\setlength{\tabcolsep}{4pt}
\renewcommand{\arraystretch}{1.15}
\begin{tabular}{l p{0.82\textwidth}}
\toprule
\textbf{Task} & \textbf{Ground-truth Dependencies (method $\rightarrow$ prerequisite methods)} \\
\midrule
ClassEval\_44 & \texttt{\_format\_line\_feed: [];\; format\_line\_html\_text: [\_format\_line\_feed];\; extract\_code\_from\_html\_text: [format\_line\_html\_text]} \\
ClassEval\_61 & \texttt{add\_song: [];\; remove\_song: [stop];\; play: [];\; stop: [];\; switch\_song: [];\; previous\_song: [];\; set\_volume: [];\; shuffle: []} \\
ClassEval\_65 & \texttt{format: [format\_string];\; format\_string: [parse\_more, trans\_three, trans\_two];\; trans\_two: [];\; trans\_three: [trans\_two];\; parse\_more: []} \\
ClassEval\_94 & \texttt{add\_item: [restock\_item];\; insert\_coin: [];\; purchase\_item: [];\; restock\_item: [];\; display\_items: []} \\
ClassEval\_43 & \texttt{add\_employee: [];\; remove\_employee: [];\; update\_employee: [get\_employee];\; get\_employee: [];\; list\_employees: []} \\
ClassEval\_84 & \texttt{read\_file\_as\_json: [read\_file];\; read\_file: [];\; write\_file: [];\; process\_file: [read\_file, write\_file]} \\
ClassEval\_91 & \texttt{add: [fix\_path];\; parse: [fix\_path];\; fix\_path: []} \\
ClassEval\_18 & \texttt{\_\_getitem\_\_: [\_convert\_key];\; \_\_setitem\_\_: [\_convert\_key];\; \_\_delitem\_\_: [\_convert\_key];\; \_\_iter\_\_: [];\; \_\_len\_\_: [];\; \_convert\_key: [\_to\_camel\_case];\; \_to\_camel\_case: []} \\
ClassEval\_34 & \texttt{read\_text: [];\; write\_text: [\_get\_alignment\_value];\; add\_heading: [];\; add\_table: [];\; \_get\_alignment\_value: []} \\
ClassEval\_36 & \texttt{send\_to: [];\; fetch: [];\; is\_full\_with\_one\_more\_email: [get\_occupied\_size];\; get\_occupied\_size: [];\; clear\_inbox: []} \\
ClassEval\_38 & \texttt{read\_excel: [];\; write\_excel: [];\; process\_excel\_data: [read\_excel, write\_excel]} \\
ClassEval\_23 & \texttt{count: [];\; count\_all: [];\; select: [\_select\_helper];\; select\_all: [select];\; \_select\_helper: []} \\
ClassEval\_58 & \texttt{generate\_mine\_sweeper\_map: [];\; generate\_playerMap: [];\; check\_won: [];\; sweep: [check\_won]} \\
ClassEval\_77 & \texttt{move: [eat\_food, reset];\; random\_food\_position: [];\; reset: [random\_food\_position];\; eat\_food: [random\_food\_position]} \\
ClassEval\_5  & \texttt{interpret: [display];\; display: []} \\
\bottomrule
\end{tabular}
\end{table*}

\subsection{Per-model Topological Order Violations}
\label{app:per_model_violations}

For each model, we list all tasks where the predicted generation order violates the ground-truth dependency constraints.
For each violated task, we report: (i) the predicted order, and (ii) at least one violated dependency edge.

\paragraph{Deepseek-V3 (4 tasks).}
\begin{itemize}
  \item \textbf{ClassEval\_44}:\\ \texttt{[\_format\_line\_feed, extract\_code\_from\_html\_text, format\_line\_html\_text]}.
  \newline Violates: \texttt{extract\_code\_from\_html\_text} $\rightarrow$ \texttt{format\_line\_html\_text}.
  \item \textbf{ClassEval\_61}:\\ \texttt{[add\_song, remove\_song, set\_volume, shuffle, play, stop, switch\_song, previous\_song]}.
  \newline Violates: \texttt{remove\_song} $\rightarrow$ \texttt{stop}.
  \item \textbf{ClassEval\_65}:\\ \texttt{[trans\_two, parse\_more, trans\_three, format, format\_string]}.
  \newline Violates: \texttt{format} $\rightarrow$ \texttt{format\_string}.
  \item \textbf{ClassEval\_94}:\\ \texttt{[add\_item, insert\_coin, purchase\_item, restock\_item, display\_items]}.
  \newline Violates: \texttt{add\_item} $\rightarrow$ \texttt{restock\_item}.
\end{itemize}

\paragraph{GPT-OSS-120B (6 tasks).}
\begin{itemize}
  \item \textbf{ClassEval\_43}: 
  \\ \texttt{[add\_employee, remove\_employee, update\_employee, get\_employee, list\_employees]}.
  \newline Violates: \texttt{update\_employee} $\rightarrow$ \texttt{get\_employee}.
  \item \textbf{ClassEval\_44}: 
  \\ \texttt{[\_format\_line\_feed, extract\_code\_from\_html\_text, format\_line\_html\_text]}.
  \newline Violates: \texttt{extract\_code\_from\_html\_text} $\rightarrow$ \texttt{format\_line\_html\_text}.
  \item \textbf{ClassEval\_61}: 
  \\ \texttt{[add\_song, remove\_song, play, stop, switch\_song, previous\_song, set\_volume, shuffle]}.
  \newline Violates: \texttt{remove\_song} $\rightarrow$ \texttt{stop}.
  \item \textbf{ClassEval\_84}:\\ \texttt{[read\_file\_as\_json, read\_file, write\_file, process\_file]}.
  \newline Violates: \texttt{read\_file\_as\_json} $\rightarrow$ \texttt{read\_file}.
  \item \textbf{ClassEval\_91}:\\ \texttt{[add, fix\_path, parse]}.
  \newline Violates: \texttt{add} $\rightarrow$ \texttt{fix\_path}.
  \item \textbf{ClassEval\_94}:\\ \texttt{[add\_item, insert\_coin, purchase\_item, restock\_item, display\_items]}.
  \newline Violates: \texttt{add\_item} $\rightarrow$ \texttt{restock\_item}.
\end{itemize}

\paragraph{Qwen2.5-Coder-7B (11 tasks).}
\begin{itemize}
  \item \textbf{ClassEval\_18}:\\ \texttt{[\_convert\_key, \_to\_camel\_case, \_\_getitem\_\_, \_\_setitem\_\_, \_\_delitem\_\_, \_\_iter\_\_, \_\_len\_\_]}.
  \newline Violates: \texttt{\_convert\_key} $\rightarrow$ \texttt{\_to\_camel\_case}.
  \item \textbf{ClassEval\_34}:\\ \texttt{[read\_text, write\_text, add\_heading, add\_table, \_get\_alignment\_value]}.
  \newline Violates: \texttt{write\_text} $\rightarrow$ \texttt{\_get\_alignment\_value}.
  \item \textbf{ClassEval\_36}:\\ \texttt{[fetch, is\_full\_with\_one\_more\_email, get\_occupied\_size, send\_to, clear\_inbox]}.
  \newline Violates: \texttt{is\_full\_with\_one\_more\_email} $\rightarrow$ \texttt{get\_occupied\_size}.
  \item \textbf{ClassEval\_38}:\\ \texttt{[read\_excel, process\_excel\_data, write\_excel]}.
  \newline Violates: \texttt{process\_excel\_data} $\rightarrow$ \texttt{write\_excel}.
  \item \textbf{ClassEval\_43}:\\ \texttt{[add\_employee, remove\_employee, update\_employee, get\_employee, list\_employees]}.
  \newline Violates: \texttt{update\_employee} $\rightarrow$ \texttt{get\_employee}.
  \item \textbf{ClassEval\_44}:\\ \texttt{[\_format\_line\_feed, extract\_code\_from\_html\_text, format\_line\_html\_text]}.
  \newline Violates: \texttt{extract\_code\_from\_html\_text} $\rightarrow$ \texttt{format\_line\_html\_text}.
  \item \textbf{ClassEval\_61}:\\ \texttt{[add\_song, remove\_song, play, stop, switch\_song, previous\_song, set\_volume, shuffle]}.
  \newline Violates: \texttt{remove\_song} $\rightarrow$ \texttt{stop}.
  \item \textbf{ClassEval\_65}:\\ \texttt{[trans\_two, trans\_three, parse\_more, format, format\_string]}.
  \newline Violates: \texttt{format} $\rightarrow$ \texttt{format\_string}.
  \item \textbf{ClassEval\_84}:\\ \texttt{[read\_file\_as\_json, read\_file, write\_file, process\_file]}.
  \newline Violates: \texttt{read\_file\_as\_json} $\rightarrow$ \texttt{read\_file}.
  \item \textbf{ClassEval\_91}:\\ \texttt{[add, parse, fix\_path]}.
  \newline Violates: \texttt{add} $\rightarrow$ \texttt{fix\_path}.
  \item \textbf{ClassEval\_94}:\\ \texttt{[add\_item, insert\_coin, restock\_item, display\_items, purchase\_item]}.
  \newline Violates: \texttt{add\_item} $\rightarrow$ \texttt{restock\_item}.
\end{itemize}

\paragraph{Qwen2.5-Coder-32B (7 tasks).}
\begin{itemize}
  \item \textbf{ClassEval\_5}:\\ \texttt{[interpret, display]}.
  \newline Violates: \texttt{interpret} $\rightarrow$ \texttt{display}.
  \item \textbf{ClassEval\_43}:\\ \texttt{[add\_employee, remove\_employee, update\_employee, get\_employee, list\_employees]}.
  \newline Violates: \texttt{update\_employee} $\rightarrow$ \texttt{get\_employee}.
  \item \textbf{ClassEval\_44}:\\ \texttt{[\_format\_line\_feed, extract\_code\_from\_html\_text, format\_line\_html\_text]}.
  \newline Violates: \texttt{extract\_code\_from\_html\_text} $\rightarrow$ \texttt{format\_line\_html\_text}.
  \item \textbf{ClassEval\_61}:\\ \texttt{[add\_song, remove\_song, play, stop, switch\_song, previous\_song, set\_volume, shuffle]}.
  \newline Violates: \texttt{remove\_song} $\rightarrow$ \texttt{stop}.
  \item \textbf{ClassEval\_84}:\\ \texttt{[read\_file\_as\_json, read\_file, write\_file, process\_file]}.
  \newline Violates: \texttt{read\_file\_as\_json} $\rightarrow$ \texttt{read\_file}.
  \item \textbf{ClassEval\_91}:\\ \texttt{[add, parse, fix\_path]}.
  \newline Violates: \texttt{add} $\rightarrow$ \texttt{fix\_path}.
  \item \textbf{ClassEval\_94}:\\ \texttt{[add\_item, insert\_coin, restock\_item, display\_items, purchase\_item]}.
  \newline Violates: \texttt{add\_item} $\rightarrow$ \texttt{restock\_item}.
\end{itemize}

\paragraph{Qwen3-Coder-30B (9 tasks).}
\begin{itemize}
  \item \textbf{ClassEval\_23}:\\ \texttt{[count, count\_all, select, select\_all, \_select\_helper]}.
  \newline Violates: \texttt{select} $\rightarrow$ \texttt{\_select\_helper}.
  \item \textbf{ClassEval\_36}:\\ \texttt{[fetch, is\_full\_with\_one\_more\_email, get\_occupied\_size, clear\_inbox, send\_to]}.
  \newline Violates: \texttt{is\_full\_with\_one\_more\_email} $\rightarrow$ \texttt{get\_occupied\_size}.
  \item \textbf{ClassEval\_43}:\\ \texttt{[add\_employee, remove\_employee, update\_employee, get\_employee, list\_employees]}.
  \newline Violates: \texttt{update\_employee} $\rightarrow$ \texttt{get\_employee}.
  \item \textbf{ClassEval\_44}:\\ \texttt{[\_format\_line\_feed, extract\_code\_from\_html\_text, format\_line\_html\_text]}.
  \newline Violates: \texttt{extract\_code\_from\_html\_text} $\rightarrow$ \texttt{format\_line\_html\_text}.
  \item \textbf{ClassEval\_61}:\\ \texttt{[add\_song, remove\_song, play, stop, switch\_song, previous\_song, set\_volume, shuffle]}.
  \newline Violates: \texttt{remove\_song} $\rightarrow$ \texttt{stop}.
  \item \textbf{ClassEval\_77}:\\ \texttt{[random\_food\_position, eat\_food, move, reset]}.
  \newline Violates: \texttt{move} $\rightarrow$ \texttt{reset}.
  \item \textbf{ClassEval\_84}:\\ \texttt{[read\_file\_as\_json, read\_file, write\_file, process\_file]}.
  \newline Violates: \texttt{read\_file\_as\_json} $\rightarrow$ \texttt{read\_file}.
  \item \textbf{ClassEval\_91}:\\ \texttt{[add, parse, fix\_path]}.
  \newline Violates: \texttt{add} $\rightarrow$ \texttt{fix\_path}.
  \item \textbf{ClassEval\_94}:\\ \texttt{[add\_item, insert\_coin, restock\_item, display\_items, purchase\_item]}.
  \newline Violates: \texttt{add\_item} $\rightarrow$ \texttt{restock\_item}.
\end{itemize}

\paragraph{Qwen3-Coder-480B (4 tasks).}
\begin{itemize}
  \item \textbf{ClassEval\_44}:\\ \texttt{[\_format\_line\_feed, extract\_code\_from\_html\_text, format\_line\_html\_text]}.
  \newline Violates: \texttt{extract\_code\_from\_html\_text} $\rightarrow$ \texttt{format\_line\_html\_text}.
  \item \textbf{ClassEval\_61}:\\ \texttt{[add\_song, remove\_song, set\_volume, shuffle, play, stop, switch\_song, previous\_song]}.
  \newline Violates: \texttt{remove\_song} $\rightarrow$ \texttt{stop}.
  \item \textbf{ClassEval\_65}:\\ \texttt{[trans\_two, parse\_more, trans\_three, format, format\_string]}.
  \newline Violates: \texttt{format} $\rightarrow$ \texttt{format\_string}.
  \item \textbf{ClassEval\_94}:\\ \texttt{[add\_item, insert\_coin, restock\_item, display\_items, purchase\_item]}.
  \newline Violates: \texttt{add\_item} $\rightarrow$ \texttt{restock\_item}.
\end{itemize}

\paragraph{Qwen3-235B (6 tasks).}
\begin{itemize}
  \item \textbf{ClassEval\_44}:\\ \texttt{[\_format\_line\_feed, extract\_code\_from\_html\_text, format\_line\_html\_text]}.
  \newline Violates: \texttt{extract\_code\_from\_html\_text} $\rightarrow$ \texttt{format\_line\_html\_text}.
  \item \textbf{ClassEval\_58}:\\ \texttt{[generate\_mine\_sweeper\_map, generate\_playerMap, sweep, check\_won]}.
  \newline Violates: \texttt{sweep} $\rightarrow$ \texttt{check\_won}.
  \item \textbf{ClassEval\_61}:\\ \texttt{[add\_song, remove\_song, set\_volume, play, stop, switch\_song, previous\_song, shuffle]}.
  \newline Violates: \texttt{remove\_song} $\rightarrow$ \texttt{stop}.
  \item \textbf{ClassEval\_65}:\\ \texttt{[trans\_two, parse\_more, trans\_three, format, format\_string]}.
  \newline Violates: \texttt{format} $\rightarrow$ \texttt{format\_string}.
  \item \textbf{ClassEval\_91}:\\ \texttt{[add, fix\_path, parse]}.
  \newline Violates: \texttt{parse} $\rightarrow$ \texttt{fix\_path}.
  \item \textbf{ClassEval\_94}:\\ \texttt{[add\_item, insert\_coin, restock\_item, purchase\_item, display\_items]}.
  \newline Violates: \texttt{add\_item} $\rightarrow$ \texttt{restock\_item}.
\end{itemize}

\paragraph{Gemini3-Flash (3 tasks).}
\begin{itemize}
  \item \textbf{ClassEval\_44}:\\ \texttt{[\_format\_line\_feed, extract\_code\_from\_html\_text, format\_line\_html\_text]}.
  \newline Violates: \texttt{extract\_code\_from\_html\_text} $\rightarrow$ \texttt{format\_line\_html\_text}.
  \item \textbf{ClassEval\_61}:\\ \texttt{[add\_song, remove\_song, set\_volume, shuffle, play, stop, switch\_song, previous\_song]}.
  \newline Violates: \texttt{remove\_song} $\rightarrow$ \texttt{stop}.
  \item \textbf{ClassEval\_94}:\\ \texttt{[add\_item, insert\_coin, restock\_item, display\_items, purchase\_item]}.
  \newline Violates: \texttt{add\_item} $\rightarrow$ \texttt{restock\_item}.
\end{itemize}

\end{document}